%% file: main2025.tex
\documentclass[11pt]{scrartcl} 

\input{structure2025.tex} 
\usepackage[normalem]{ulem}

\title{\Large{Modeling phononic band gap in microstructured solids using the
Riemann-Cartan geometric framework 
}}

\author[1]{\large Ilya Peshkov}
\author[2]{Lo\"ic Le Marrec}
\affil[1]{\textit{\small University of Trento, DICAM, Trento, Italy}}
\affil[2]{\textit{\small Univ Rennes, CNRS, IRMAR - UMR 6625, F-35000 Rennes, France}}

\date{\small\today} 

\begin{document}

    \maketitle
	\begin{abstract}\noindent\small This paper discusses the modeling of
	acoustic wave fields in microstructured elastic solids within the framework
	of Riemann-Cartan geometry. We consider a scenario in which microstructural
	deformations occur significantly faster than those of the bulk material.
	This time-scale separation creates apparent geometric incompatibilities at
	the macroscopic level, even in the absence of permanent inelastic
	deformation or damage. We formalize this phenomenon by using a non-holonomic
	frame field to represent macroscopic elastic deformations and an associated
	torsion field to characterize the resulting geometric incompatibilities. The
	spatial components of the torsion tensor quantify the instantaneous
	geometric incompatibility of the macroscopic deformations, while its
	temporal components capture the inertial effects arising from the reversible
	energy exchange between the micro- and macro-scales. A key finding is that
	the model's dispersion relation predicts a complete frequency band gap.
	Furthermore, the governing equations exhibit a mathematical analogy to
	Maxwell's equations, potentially bridging the modeling of phononic and
	photonic metamaterials.
	\end{abstract}

\section{Introduction}

We discuss the modeling of acoustic wave fields in microstructured elastic
solids. The microstructure might be of different nature, such as continuously
distributed dislocations in crystalline solids, or periodic inclusions and
resonators in artificially designed acoustic/phononic metamaterials. We are
particularly interested in the conditions when the microstructural features can
significantly alter the propagation of acoustic waves due to a reversible
coupling of the micro and macro scales \cite{Nabarro1951,Cummer2016}. 

To deal with such scenarios, we aim at developing a system of time-dependent
partial differential equations (PDEs) that can capture the essential physics of
wave propagation in microstructured solids without resolving the details of the
microstructure itself. This approach is particularly useful when the
microstructural features are much smaller than the wavelength of interest,
allowing us to treat the medium as effectively homogeneous at the macroscopic
scale.

In this paper, we propose such a PDE system in the framework of the
four-dimensional (4D) Riemann-Cartan geometry. The main fields of the
model are two 4D frame fields, one of which ($\Dist_A$, $A=0,1,2,3$) is used to
represent the macroscopic, or observable, deformations which are associated with
a length scale $L$ representing a resolution limit of our experimental or
computational tools. The second frame field ($\Plast_a$, $a=0,1,2,3$) is used to
capture the deformation of the microstructural elements, which are associated to
a certain length scale $\ell \ll L$ and are assumed to be unavailable to our
direct observations.

Despite we are interested only in the elastic/reversible deformations of the
material (i.e. no plasticity, damage, phase change, etc), the macroscopic frame
field $\Dist$ is non-holonomic in general (i.e. it cannot be represented as a
gradient of a vector field), meaning that it describes non-compatible
deformations at the macroscopic level. This non-holonomic nature arises from the
fact that the microstructural deformations can occur on a much faster time scale
than the macroscopic ones, leading to apparent geometric incompatibilities at
the macroscopic scale. However, these incompatibilities are only instantaneous
(or dynamical) and do not lead to any permanent inelastic deformation or damage.
This means that if all the vibrations are damped and the material is brought at
equilibrium, the macroscopic frame field becomes compatible and the solid
material restores its shape.    

The geometric incompatibility of the macroscopic elastic deformations is
represented by the torsion tensor of the non-holonomic frame field. The spatial
components of the 4D torsion tensor (the spatial curl of the frame field)
quantify the instantaneous geometric incompatibility, while its time components
capture the inertial effects arising from the reversible energy exchange between
the micro and macro scales. Importantly, the introduction of the torsion field
results in the appearance of additional propagating modes, which can be
naturally interpreted as modes associated to the relative rotations (due to the
curl operator) of the macroscopic and microscopic frame fields.

The basic underlying idea of the proposed approach consisting in using two frame
fields to represent the macro and micro scales is not new and can be tracked
back to the works
\cite{Cosserat1909,Nye1953,Bilby1955,Toupin1962,Kroner1963,Mindlin1964,
Kosevich1965,Eringen1964a,Madeo2015a}, and others. Likewise, for long time there
is a consensus on the fact that one needs to use higher gradients of the
deformation field to capture the material heterogeneity
\cite{DellIsola2017,DellIsola2022}. However, particular realization details are
different from model to model. In particular it is not clear in what form the
higher gradients should enter the governing equations, how many additional
fields are needed to represent the microstructure, how the micro and macro
scales should be coupled, etc. For example, a quite general linear model was
suggested in \cite{Mindlin1964} and nonlinear model in \cite{Eringen1964a} with
a large number of parameters. These models exploit the full gradient of the
deformation field. However, as discussed in \cite{Madeo2015a}, one can already
obtain a quite general model for microstructured solids if only the curl (not
the full gradient) of the deformation field is used.

The curl of the frame field used in \cite{Madeo2015a} can be seen as the spatial
part of the four dimensional torsion tensor. Therefore, motivated by results in
\cite{Madeo2015a}, we develop our model for microstructured solids in the
Riemann-Cartan geometric framework with non-zero torsion and zero curvature. The
resulting model is nonlinear and can be applied not only to modeling propagation
of waves with small amplitude but also to modeling of large elastic deformations
of microstructured solids. However, in this paper we restrict our consideration
to small deformations, and in particular we show that the model predicts a
frequency band gap, a typical phenomenon for phononic metamaterials.

We note that the idea of modeling microstructured solids using a Riemann-Cartan
geometry is also not new, e.g.
\cite{Cartan1922,Hehl2007,LychevKoifman2018,NguyenLeMarrec2022,CRESPO2025a}, but
it is more commonly associated with modeling of continuously distributed
dislocations \cite{Kleinert1989,lazar2000,Lazar2002a,MALYSHEV2007}. 

The system of governing equations is formulated as a system of first-order
hyperbolic equations with relaxation source terms. Importantly, the relaxation
terms are of reversible nature, i.e they do not rise the entropy but describe
the reversible energy exchange between the macro and micro scales. We note that
hyperbolic equations is a standard tool for modeling wave propagation in
continuous media. However, the presence of relaxation source terms may
dramatically affect how waves propagate in the medium. We analyze the dispersion
relation of the proposed model and show that such a system of first-order
hyperbolic equations may exhibit non-propagating modes (standing waves) -- the
frequency band gap. This is not typical for first order hyperbolic equations. At
least, we are aware of only one other example of first-order hyperbolic PDEs
exhibiting a band-gap \cite{Lombard2019}.

The novelty of our formulation is multifold. First, we define the torsion tensor
from the macroscopic frame field but not from the microstructural one as it is
done for example in \cite{Madeo2015a}. Second, we formulate the model in the
four-dimensional formalism, such that the time component of the torsion field
plays an important role in representing the microinertial effects arising from
the reversible energy exchange between the micro and macro scales. Third, we
formulate the governing equations in a thermodynamically compatible way,
ensuring that they are consistent with the principles of thermodynamics. Fourth,
the system of governing equations is formulated as a system of first-order
hyperbolic equations. Finally, we show that the governing equations exhibit a
notable mathematical analogy to Maxwell's equations, which opens up new avenues
for understanding and designing phononic metamaterials using the analogies with
photonic metamaterials.

The paper is organized as follows. In Section\,\ref{sec.def}, we introduce the
basic definitions and notations used throughout the paper. In
Section\,\ref{sec.model}, we present the governing equations of the proposed
model and discuss their thermodynamic consistency. In Section\,\ref{sec.DR}, we
analyze the dispersion relation of the model and demonstrate the existence of a
frequency band gap. Finally, in Section\,\ref{sec.conclusion}, we summarize our
findings and discuss potential future research directions.

\section{Definitions and notations}\label{sec.def}

To appreciate the differential geometric structure of the theory of
microstructured solids, we need to follow the four-dimensional formalism at
least at the beginning of the paper when the basic definitions are introduced.
The final governing differential equations, however, will be presented in the
usual three-dimensional form.

We use the following index conventions: Greek indices $ \mu, \nu, \ldots =
0,1,2,3 $ denote the four-dimensional space-time indices, Latin indices $
i,j,k,\ldots = 1,2,3 $ denote the three-dimensional spatial indices, while Latin
indices $ A,B,\ldots = 0,1,2,3 $ denote the Lagrangian (mater-time) indices, and
upper case mid-alphabet Latin indices $I,J,K,\ldots$ denote the Lagrangian
material indices. Moreover, $a,b,c,\ldots=0,1,2,3$ will be used to denote the
space-time components of tensors in the local frames associated with the
microelements. The Einstein summation convention is used throughout the paper.

Let $x^\mu=\left\{ x^0,x^k \right\}$, $\mu=0,1,2,3$ be the space-time Cartesian
coordinate system (e.g. associated with a fixed laboratory frame
$\bm{\pd}_\mu:=\frac{\pd}{\pd x^\mu}$, assumed to be orthonormal) with $x^0=t$
being the time coordinate. 

In the standard approach to continuum mechanics, if $X^A$ are the space-time
coordinates of the elastic body in its relaxed (stress-free) configuration, then
the motion of the medium is described by a one-to-one mapping $x^\mu =
\hat{x}^\mu(X^A),$ between the coordinates $X^A$ of the medium's reference
configuration and coordinates $x^\mu$ of its current configuration. The
deformation of the material is then described by the deformation gradient
\begin{equation}\label{eq.F}
	F^\mu_{\ A}:=\frac{\pd \hat{x}^\mu}{\pd X^A}.
\end{equation}
which also can be seen as the components of the coordinate frame field
$\bm{\pd}_A:=\frac{\pd}{\pd X^A}$ in the laboratory frame $\bm{\pd}_\mu$:
\begin{equation}\label{eq.frame}
	F^\mu_{\ A} \bm{\pd}_\mu = \bm{\pd}_A.
\end{equation}

More explicitly, the four-deformation gradient can be written as
\begin{equation}
	F^\mu_{\ A} = \begin{pmatrix}
		1   & 0 & 0 		& 0 \\
		v^1 &   &   		& 	\\
		v^2 &   & F^k_{\ I} & 	\\
		v^3 &   & 			&	
	\end{pmatrix}
\end{equation}
where 
\begin{equation}
	v^\mu = (1,v^k) := F^\mu_{\ 0} = \frac{\pd \hat{x}^\mu}{\pd t}
\end{equation}
is the velocity of the material element w.r.t. the laboratory frame, and $F^k_{\
I} = \pd \hat{x}^k/\pd X^I$ is the spatial part of the four-deformation
gradient.

Although, we assume that the material does not undergo any permanent, i.e.
\textit{irreversible}, deformations, we can not use the standard approach and
the existence of the one-to-one mapping \eqref{eq.F} because due to the presence
of the microstructure, the material microelements can deform on a much faster
time scale than the macroelements resulting in apparent geometric
incompatibilities at the macroscale. Instead, we will work directly in terms of
the local frame fields, i.e. with the relation \eqref{eq.frame}, which in
general will be assumed to be non-holonomic (i.e. they are not coordinate basis
of a coordinate system in general, except the laboratory frame $\bm{\pd}_\mu$
which always assumed to be a coordinate basis of $x^\mu$).

Thus, we assume the macroelements of the material are associated with the local
frame field $\bm{h}_A$ which is co-moving and co-deforming with the elements
such that at each time instant $t$ its coordinates in the laboratory frame are
given by $F^\mu_{\ A}$, i.e.
\begin{equation}
	\bm{h}_A =  F^\mu_{\ A} \bm{\pd}_\mu.
\end{equation}
We shall associate the frame field $\bm{h}_A$ with certain time $T$ and length
$L$ scales, called macro scales, at which the dynamics of the continuum can be
observed by a laboratory observer who has tools which can measure (observe) the
material at the scales $T$ and $L$.

Furthermore, following the ideas of Eringen \cite{Eringen1964a,Eringen1968}, we
assume that the macroelements themselves are composed of microelements which can
deform independently from the macroelements. The microelements are associated
with another local frame field $\bm{\eta}_a$ whose coordinates in the material
frame $\bm{h}_A$ are denoted as $F^A_{\ a}$, i.e.
\begin{equation}
	\bm{\eta}_a = F^A_{\ a} \bm{h}_A.
\end{equation} 
The micro frame field $\bm{\eta}_a$ is associated with certain time $\tau$ and length $\ell$ scales, called micro scales, such that $\tau \ll T$ and $\ell \ll L$.

In the following, we shall also refer to the matrices $F^\mu_{\ A}$ and $F^A_{\
a}$ as macro and micro \textit{distortions}.

Note that from the point of view of a microscopic observer\footnote{A
microscopic observer is a laboratory observer who has tools with resolution up
to the microscopic time and length scales $\tau$ and $\ell$.}, the deformation
of the continuum is fully compatible, i.e. the micro frame field $\bm{\eta}_a$
is holonomic and can be associate with the coordinate basis
$\bm{\pd}_a:=\frac{\pd}{\pd \xi^a}$ of a certain coordinate system $\xi^a$, and
the coordinates of the micro frame field $\bm{\eta}_a$ in the laboratory frame
$\bm{\pd}_\mu$ are given by
\begin{equation}
	F^\mu_{\ a} = F^\mu_{\ A} F^A_{\ a},
\end{equation}  
which can be seen as the decomposition of the total four-deformation gradient
$F^\mu_{\ a}$ into the product of the macro and micro distortions.  
Note that the geometrical compatibility of the micro deformations means that 
\begin{equation}\label{eq.rotF}
	\pd_b F^\mu_{\ a} - \pd_a F^\mu_{\ b} = 0.
\end{equation}
Essentially, this implies that from the microscopic point of view, there is a
one-to-one mapping $x^\mu = \tilde{x}^\mu(\xi^a)$ and $F^\mu_{\ a} = \frac{\pd
\tilde{x}^\mu}{\pd \xi^a}$ and \eqref{eq.rotF} simply expresses the commutation
of the partial derivatives, $\pd_a\pd_b \tilde{x}^\mu - \pd_b\pd_a \tilde{x}^\mu
= 0$.

However, in this paper, we exclude the microscopic observer's point of view. The
fine time and length scales $(\tau, \ell)$ are considered unobservable. Instead,
we are restricted to macroscopic time and length scales $T\gg \tau$ and $L \gg
\ell$. This scale separation implies two qualitatively different dynamic
regimes. Specifically, if the characteristic time and length scale of the
perturbations are on the order of $T$ and $L$, the microstructure's dynamics is
not activated, and the material effectively behaves as a classical, single-scale
material. In this case the macro and micro frames become equivalent,
$\bm{\eta}_a \sim \bm{h}_A$, and hence the macro distortion $F^\mu_{\ A}$ is
holonomic. However, if the characteristic time and length scale of the waves are
sufficiently smaller than $T$ and $L$, then the microstructure's dynamics can be
activated resulting in temporal geometrical incompatibilities at the macroscale,
i.e. the macro distortion $F^\mu_{\ A}$ becomes non-holonomic, i.e. an equality
similar to \eqref{eq.rotF} does not hold for $F^\mu_{\ A}$. 

We emphasize that the geometrical incompatibilities observed at the macro scale is
strictly an emergent phenomenon, resulting only from the coarse-graining of
observations that do not resolve the microstructure's details. Fundamentally,
the same underlying dynamics are geometrically compatible and holonomic when
viewed at the micro scale. In essence, increasing the resolution of our
observational tools eliminates the perceived incompatibility.

From the geometrical perspective, the local incompatibility of the macroscopic
deformations means that the instantaneous geometry of the medium is not
Euclidean, i.e. it cannot be represented by a flat material manifold. To
describe this, one needs to use the tools of differential geometry. The well
known measure of non-flatness of a geometry is the Riemann curvature tensor,
which can be expressed in terms of second derivatives of the frame field.
However, another less known measure of non-flatness is the torsion tensor, which
is made of first-order derivatives of the frame field. Since we want to keep the
model as simple as possible, it is natural to try to work in terms of the
torsion field first. 

Therefore, in this paper, to capture the effect of the microstructure without
resolving its details, we shall model the dynamics of the geometrical
incompatibilities at the macroscale by introducing the torsion
tensor of the macro distortion
\begin{equation}
	\Tors{A}{\mu\nu} := \pd_\mu A^A_{\ \nu} - \pd_\nu A^A_{\ \mu},
\end{equation}
where $A^A_{\ \mu}$ is the inverse of $F^\mu_{\ A}$, i.e.
\begin{equation}
	A^B_{\ \mu} F^\mu_{\ A} = \delta^B_{\ A}, \qquad F^\mu_{\ B} A^B_{\ \nu} = 
	\delta^\mu_{\ \nu},
\end{equation}
with $\delta^B_{\ A}$ and $\delta^\mu_{\ \nu}$ being the Kronecker deltas.

The four-torsion tensor $\Tors{A}{\mu\nu}$ can be decomposed into its temporal
and spatial parts as
\begin{equation}
	E^A_{\ \mu} := \Tors{A}{\mu \nu} v^\nu, \qquad 
	B^{A\mu} := \HT{A\mu\nu} v_\nu,
\end{equation}
where 
\begin{equation}
	\HT{A\mu\nu} 
	:= \frac12 \varepsilon^{\mu\nu\alpha\beta} \Tors{A}{\alpha\beta}
\end{equation}
is the Hodge dual of the torsion tensor, $\varepsilon^{\mu\nu\alpha\beta} $ is
the four-dimensional Levi-Civita symbol, $ v_\mu = G_{\mu\nu} v^\nu =
(1,0,0,0)$, and 
\begin{equation}
	G_{\mu\nu} := A^A_{\ \mu} \eta_{AB} A^B_{\ \nu}  
\end{equation}
 is the material metric tensor in the laboratory frame $\bm{\pd}_\mu$, while
 $\eta_{AB}=\text{diag}(1,1,1,1)$ is the material metric in the macroscopic
 reference frame $\bm{h}_A$. Later, we shall also need the space time metric
 $\eta_{\mu\nu} = \text{diag}(1,1,1,1)$ and the microscopic metric $\eta_{ab} =
 \text{diag}(1,1,1,1)$. In general, the macroscopic metric $\eta_{AB}$ and
 microscopic metric $\eta_{ab}$ can be anisotropic to account for the anisotropy
 of the material. However, in this paper we restrict ourselves to isotropic
 materials for simplicity. The space time metric $\eta_{\mu\nu}$ is always
 assumed to be isotropic. 

 Then, assuming a general form of the Lagrangian density
\begin{equation}\label{eqn.Lagrangian}
	\mathcal{L} = \mathcal{L}(A^A_{\ \mu}, E^A_{\ \mu}, B^{A\mu}),
\end{equation}
and after a proper splitting of the four tensors on their temporal and spatial
part, one can derive the governing equations using the variational principle. We
omit the detailed derivation here for brevity, and present the final system of
governing equations in the next section. The interested reader can find the
details in \cite{Peshkov2025}, while a more heuristic derivation can be also
found in \cite{PRD-Torsion2019}. However, in this way, we obtain a system of
PDEs only for the fields $A^A_{\ \mu}$, $E^A_{\ \mu}$, and $B^{A\mu}$ but not
for the microscopic frame field. To close the system, we will assume a very
simple evolution equation for the microdistortion $F^A_{\ a}$ (actually its
inverse which will be denoted $P^a_{\ A}$). This will be discussed later.

Before proceeding, we make an important remark on the use of the
four-dimensional formalism and analogies between the proposed model and
Maxwell's equations. First of all, we argue in favor of the four-dimensional	
formulation because it allows to naturally introduce the microinertial effects
via the time components $E^A_{\ \mu}$ of the torsion tensor, which are usually
overlooked in 3D models of microstructured solids, e.g. models for dynamics of
continuously distributed dislocations
\cite{Dzyaloshinskii1980,Katanaev2005,clayton2006,Acharya2022}. Second, the
four-dimensional formalism reveals a intriguing mathematical analogies between the
governing equations of our model (discussed in the next section) and Maxwell's
equations of electromagnetism. This analogy may help in understanding and
designing phononic metamaterials using concepts from photonic metamaterials.

\section{The model}\label{sec.model}

\subsection{Governing equations}

In this section, we discuss the governing equations of the proposed model for
microstructured solids. A detailed derivation of these equations, based on the
variational principle, can be found in \cite{Peshkov2025}. Although the
derivation in \cite{Peshkov2025} was performed in the relativistic context, the
resulting equations are broadly applicable to other physical systems, such as
microstructured solids, due to the universality of differential geometry
concepts. Furthermore, while \cite{Peshkov2025} employs two distinct frame
fields (space-time and matter frame) with potentially different velocities, we
adopt a simpler setting here: we assume that both the macro and micro frame
fields are co-moving with the same velocity $v^i$. Importantly, their
deformations remain independent. The assumption of a single velocity field
reduces the complexity of the governing equations relative to the full
multi-velocity case. This formulation allows for the microdistortion field to be
governed by a straightforward transport equation, and the advective terms in all
the governing equations are mediated exclusively by the macroscopic velocity
$v^i$.

We note that in this paper we only focus on the governing equations for
microstructured solids, while the question of boundary conditions is not
addressed here. Thus, we shall assume that the medium is infinite and
homogeneous. The treatment of boundary conditions, in particular for interfaces
between microstructured and simple solids, will be addressed in future works. 

In what follows, we will slightly abuse index notations, and we will use both
capital Latin indices $ A,B,\ldots = 1,2,3 $ and lower case Latin indices $
a,b,\ldots=1,2,3$ to denote only space components of tensors (not space-time
components as in the previous section) in the local frames $\bm{h}_A$ and
$\bm{\eta}_a$ associated with the macro and micro elements respectively.

The final system of governing PDEs of the proposed model reads (this is
essentially system \cite[eq.(31)]{PRD-Torsion2019} or system
\cite[eq.(79)]{Peshkov2025} enriched with the evolution equation for the
microdistortion $P^a_{\ A}$):
\begin{subequations}\label{eqn.PDE}
	\begin{align}
	&\pd_t \durg{A}{i} + \pd_k \left( \durg{A}{i} 
	\vel{k}  -  \vel{i} 
	\durg{A}{k} - \LeviCivitaUp{i k j}  H_{jA} \right) + \vel{i} 
	\pd_k\durg{A}{k}  = \frac{1}{\alpha}\Pi^i_{\ A} + \frac{1}{\beta}\pi^i_{\ A}, 
		\label{eqn.PDE.extend.D}\\[2mm]
	&\pd_t \burg{A}{i} + \pd_k \left(
	\burg{A}{i} \vel{k} - \vel{i} \burg{A}{k} + \LeviCivitaUp{ikj} 
	E^A_{\ j}
	\right) + \vel{i} \pd_k \burg{A}{k} = 0,\label{eqn.PDE.extend.B}
	\\[2mm]
	&\pd_t M_i + \pd_k \left( M_i \vel{k} - \Sigma^k_{\ i} 
	\right) = 0,\label{eqn.PDE.extend.M}
	\\[2mm]
	&\pd_t \dist{A}{k} + \vel{j}\pd_j \dist{A}{k} + \dist{A}{j}\pd_k \vel{j}
	 = 
	-\frac{1}{\alpha} E^A_{\ k},\label{eqn.PDE.extend.A}
	\\[2mm]
	&\pd_t P^a_{\ A} + v^k \pd_k P^a_{\ A} = -\frac{1}{\beta} E^a_{\ A},
	\label{eqn.PDE.extend.P}
	\end{align}
\end{subequations}
where we used the following notations:
\begin{subequations}\label{eqn.therm.forces}
	\begin{align}
		\Pi^k_{\ A} := \frac{\pd \En}{\pd \dist{A}{k}}, 
		\qquad 
		& \pi^A_{\ a} := \frac{\pd \En}{\pd P^a_{\ A}},\\[2mm]
		E^{A}_{\ i} := \frac{\pd \En}{\pd \durg{A}{i}}, 
		\qquad 
		& H_{iA} := \frac{\pd \En}{\pd \burg{A}{i}},
	\end{align}
\end{subequations}
which are called thermodynamic forces conjugated to the fields $ \dist{A}{k} $,
$ P^A_{\ i} $, $\durg{A}{i}$, and $\burg{A}{i}$ respectively, while $\En =
\En(M_i,A^A_{\ k},\allowbreak \durg{k}{A},\burg{A}{k},P^a_{\ A})$ is the total energy of the
system. Furthermore, sometimes we need to write these quantities in a different
frame, e.g.  $\pi^i_{\ A}$ and $E^a_{\ A}$ are the thermodynamic forces defined
in \eqref{eqn.therm.forces} but written in different frames, i.e.
\begin{equation}
	\pi^i_{\ A} := \pi^B_{\ a} F^i_{\ B} P^a_{\ A}, 
	\qquad 
	E^a_{\ A} := E^{B}_{\ i} F^i_{\ A} P^a_{\ B}.
\end{equation}  

The momentum $M_i$ is defined as the total momentum of the system which
consists of two parts: the translation momentum of the medium $m_i = \rho v_i$
and the momentum associated with the torsion fields, i.e.
\begin{equation}
	M_i = \rho v_i + \varepsilon_{ijk} \durg{A}{j} \burg{A}{k}.
\end{equation}
or in vector calculus notation
\begin{equation}
	\MM = \rho \vv + \bm{D}_A \times \bm{B}^A,
\end{equation}
where $\bm{D}_A$ and $\bm{B}^A$ are rows and columns of the matrices
$\durg{A}{i}$ and $\burg{A}{i}$ respectively, and $\varepsilon_{ijk}$ is the 3D
Levi-Civita symbol. In our approach, which can be also seen as a Hamiltonian
formulation \cite{SHTC-GENERIC}, the total momentum $M_i$ is a primary
variable, while the velocity $v_i$ is defined from the energy potential $\En$ as
a conjugated variable, i.e. $v^i = \pd \En/\pd M_i$. Also, $ \rho $ is the
mass density which is computed as $ \rho = \rho_0 \det(\dist{A}{k}) $ with $
\rho_0 $ being the reference mass density.

The relaxation parameters $ \alpha > 0 $ and $ \beta > 0 $ in
\eqref{eqn.PDE.extend.D} and \eqref{eqn.PDE.extend.P}
have the units of inverse length, $\ell^{-1}$. They control the rate of energy
exchange between the micro and macro scales. In this paper, we assume that they
are constant in time and uniform in space.

Note that the precise definition of the 3D $\bm{B}=\left\{ \burg{A}{i} \right\}$
field is $\burg{A}{i} = \alpha\varepsilon^{ijk} \pd_j \dist{A}{k}$, see
\cite{PRD-Torsion2019}, while the definition of the 3D $\bm{D} = \left\{
\durg{A}{i} \right\}$ field is more involved and it cannot be simply expressed
in terms of the space-time gradient of the distortion field $\dist{A}{k}$. More
precisely, $\durg{A}{i}$ is introduced as a thermodynamically dual field to
$E^A_{\ i}$, i.e. $\durg{A}{i} = \frac{\pd \mathcal{L}}{\pd E^A_{\ i}}$, see
\cite{Peshkov2025}, exactly as electric displacement is thermodynamically dual
to the electric field in nonlinear electrodynamics
\cite{Eringen1991,EringenMauginII}. Here, $\mathcal{L}$ is the Lagrangian
density of the system \eqref{eqn.Lagrangian}. Essentially, $\durg{A}{i}$
inherits from $E^A_{\ i}$ the relation to the time derivative of the distortion
field $\dist{A}{k}$, see \eqref{eqn.PDE.extend.A}, in the same way as electric
displacement $\bm{D}$ inherits from the electric field $\bm{E}$ the relation to
the time derivative of the vector potential in electrodynamics. Thus,
$\bm{D}$ carries the information about the inertial effects associated with the
time variations of the distortion field $\dist{A}{k}$.

Furthermore, in \eqref{eqn.PDE.extend.M},
\begin{equation}\label{stress}
	\Sigma^k_{\ i} := - P \kronecker{k}{i}
	- \dist{A}{i} \Pi^k_{\ A} 
	+ \durg{A}{k} E^{A}_{\ i}
	+ \burg{A}{k} H_{kA} 
\end{equation}
is the total stress tensor, $ P := M_i v^i + \durg{A}{i} E^{A}_{\ i} +
\burg{A}{i} H_{iA} - \En $ is the thermodynamic pressure. 

Let us make a remark about the time evolution of the microdistortion field
$P^a_{\ A}$ governed by \eqref{eqn.PDE.extend.P}. This equation is not derived
from the variational principle as the other equations in \eqref{eqn.PDE}, but is
postulated in order to close the system. As one can see this PDE is rather
simple, i.e. it is a pure transport equation with a source term. The rationale
behind this choice is multifold. First, from the macroscopic observer's point of
view, the matrix $P^a_{\ A}$ is not a tensor field but rather a collection of
scalar fields (its components) because it does not have spatial indices $i,j,k$
and therefore its entries transform as scalars under the coordinate
transformation $x^k \to x^{k'}$. Hence, the left hand side of
\eqref{eqn.PDE.extend.P} is simply the material derivative of the field $P^a_{\
A}$. Second, to allow changes in the microdistortion field, we introduce a
source term proportional to the time component of the torsion field, $E^a_{\
A}$, i.e. it is proportional to the time variation of the macroscopic distortion
field. This is quite natural from the multiscale thermodynamics point of view
because the micro and macro distortions live at different time and length
scales, and therefore can interact only via nonlocal terms, i.e. first and
higher gradients of the macroscopic fields, e.g. via the torsion field in our
case.

Furthermore, note that the microdistortion field $P^a_{\ A}$ is not contributing
to the stress tensor \eqref{stress}. However, the associated microstress
$\pi^i_{\ A}$ enters the evolution equation \eqref{eqn.PDE.extend.D} of the
torsion field $\durg{A}{i}$ as a source term. Likewise, the torsion field
$E^{A}_{\ i}$ feeds the macro distortion $A^A_{\ k}$ through the source term in
\eqref{eqn.PDE.extend.A}, while the associated macrostress $\Pi^k_{\ A}$ feeds
$\durg{A}{i}$ via the source term in \eqref{eqn.PDE.extend.D}. However, unlike
the microdistortion, the macrodistortion is also subjected to the deformation by
the velocity field via the term $A^a_{\ j}\pd_k v^j$. This is how the coupling
between the macro and micro scales is achieved in the model.

\subsection{Energy conservation}

On the solutions of \eqref{eqn.PDE}, an additional conservation law is satisfied
\begin{equation}\label{energy.cons}
	\pd_t \En + \pd_k \left( \En \vel{k} - \vel{i} \Sigma^k_{\ i} + 
	\varepsilon^{kij} E^A_{\ i}H_{jA} \right) = 0,
\end{equation}
which represents the conservation law for the total energy of the system. It can
be obtained by multiplying every equation in \eqref{eqn.PDE} by the corresponding
thermodynamic force defined in \eqref{eqn.therm.forces} and summing up all the
resulting equations:
\begin{equation}\label{eqn.sum}
	\eqref{energy.cons} = E^{A}_{\ i} \cdot \eqref{eqn.PDE.extend.D} + H_{iA} \cdot \eqref{eqn.PDE.extend.B} + v^i \cdot \eqref{eqn.PDE.extend.M} + \Pi^k_{\ A} \cdot \eqref{eqn.PDE.extend.A} + \pi^A_{\ a} \cdot \eqref{eqn.PDE.extend.P}.
\end{equation}  

Importantly, that after the summation procedure \eqref{eqn.sum}, the source
terms in \eqref{eqn.PDE} cancel out:
\begin{equation}\label{eqn.sum.source}
	E^{A}_{\ i} \left( \frac{1}{\alpha}\Pi^i_{\ A} + \frac{1}{\beta} \pi^i_{\ A} \right)
	- \Pi^k_{\ A} \left( \frac{1}{\alpha} E^{A}_{\ k} \right) - \pi^A_{\ a} \left( \frac{1}{\beta} E^a_{\ A} \right)
	\equiv 0,
\end{equation}
meaning that
they do not contribute to the total energy balance. This shows that these source
terms are of reversible nature, i.e. they do not rise the entropy but describe
the reversible energy exchange between the macro and micro scales.

\subsection{Energy potential}\label{sec.energy.cons}

The governing equations \eqref{eqn.PDE} are not closed until the energy density potential $
\En $ is specified because all the constitutive fluxes and source terms are defined
in terms of the thermodynamic forces \eqref{eqn.therm.forces} which are in turn
defined as derivatives of $ \En $. In this paper, we assume the following form
of the energy potential

\begin{equation}\label{eqn.En}
	\En(\MM,\AA,\Burg,\Durg,\bm{P}) 
	= \En^\text{macro} 
	+ \En^\text{meso} 
	+ \En^\text{micro} 
	+ \En^\text{kin}
\end{equation}
where $\En^\text{macro}$ and $\En^\text{micro}$ are the elastic energies
associated with the deformations of the bulk material and of the microstructure,
and are defined as
\begin{equation}\label{eqn.elast.macro}
	\En^\text{macro} = \En_1 + \En_2 := \frac{\rho C_0^2}{\Gamma(\Gamma - 1)} \det(\AA)^{\Gamma-1} 
	+ \frac{\rho C_s^2}{4} \Vert \bm{G}' \Vert^2,
\end{equation}
\begin{equation}\label{eqn.elast.micro}
	\En^\text{micro} = \En_3 + \En_4 := \frac{\rho c_0^2}{\gamma(\gamma - 1)} \det(\bm{P})^{\gamma-1} 
	+ \frac{\rho c_s^2}{4} \Vert \bm{g}' \Vert^2,
\end{equation}
with $C_0$ and $c_0$ being the bulk sound speeds of the macroscopic material and
microstructure respectively, while $C_s$ and $c_s$ are corresponding shear sound
speeds. Also, $ \Gamma $ and $ \gamma $ are adiabatic indices of the macroscopic
material and microstructure respectively, $\bm{G}' = \bm{G} -
\frac{1}{3}\text{tr}(\bm{G})$ is the deviatoric part of the macroscopic metric
tensor $ \bm{G} = \{G_{ij}\} $, $ G_{ij} = \dist{A}{i} \eta_{AB} \dist{B}{j} $,
and $ \bm{g}' = \bm{g} - \frac{1}{3}\text{tr}(\bm{g}) $ is the deviatoric part
of the microscopic metric tensor $ \bm{g} = \{ g_{AB} \} $, $ g_{AB} = P^a_{\ A}
\eta_{ab} P^b_{\ B} $. Here, $ \eta_{AB} = \text{diag}(1,1,1) $ and $\eta_{ab}
= \text{diag}(1,1,1) $ are the metrics in the local frames $\bm{h}_A$
and $\bm{\eta}_a$ respectively. They can be anisotropic in general but in this
paper we assume isotropy for simplicity. 

The term $\En^\text{meso}$ is the energy associated with the torsion fields, and
it defines the coupling between the macro and micro scales. In this paper, we
assume the following form of $ \En^\text{meso} $
\begin{equation}\label{energy.torsion}
\En^\text{meso}  = \En_5 + \En_6 := \frac{1}{2}\left (\frac1\epsilon \, 
\Vert\Durg\Vert^2 
+ 
\frac1\mu \, \Vert\Burg\Vert^2\right ) 
+
\frac{1}{\rho} \varepsilon_{ijk} M^i \burg{A}{j}\durg{A}{k}
\end{equation}
where, $ \epsilon $ and $ \mu $ are torsion-related transport parameters which
together scales as $ (\epsilon\mu)^{-1}\sim \text{velocity}^2 $. As we shall see
in the dispersion analysis, these two parameters introduce a velocity
\begin{equation}\label{light.speed}
\cinf = \frac{1}{\sqrt{\epsilon\mu}}
\end{equation}
which bounds from above the group velocities of the so-called optical modes in
the medium. 

Finally, the term $\En^\text{kin}$ is the kinetic energy of the
medium defined as
\begin{equation}
	\En^\text{kin} = \En_7 := \frac{1}{2\rho} \Vert \MM \Vert^2,
\end{equation}
which contains contribution from both the translational momentum of the medium
and from the torsion fields.

\subsection{Symmetric hyperbolicity}\label{sec.symmetric}

The system of governing equations \eqref{eqn.PDE} is a nonlinear system, and one
has to make sure that it is well posed. In fact, it can be shown that the system
\eqref{eqn.PDE} can be transformed into an equivalent symmetric hyperbolic form
\cite{God1961,Rom1998,Rom2001,GodRom2003,SHTC-GENERIC}:
\begin{subequations}
	\begin{gather}
		\mathbb{M}\frac{\pd \bm{p}}{\pd t} 
		+ \mathbb{N}^k\frac{\pd \bm{p}}{\pd x^k} 
		= \bm{S}(\bm{p}),\\
		\mathbb{M}^\transpose = \mathbb{M}>0, 
		\qquad (\mathbb{N}^k)^\transpose = \mathbb{N}^k,
	\end{gather}
\end{subequations}
provided that the energy potential $ \En $ is convex with respect to the state
variables. Here, $\mathbb{M}(\bm{p})$ and $\mathbb{N}^k(\bm{p})$ are coefficient
matrices, and $\bm{S}(\bm{p})$ is the vector of all algebraic source terms in
\eqref{eqn.PDE}. 

As it is known from the theory of symmetric hyperbolic systems
\cite{Kato1975,Serre2007}, such systems has local-in-time existence and
uniqueness of the solution. In fact, system \eqref{eqn.PDE} falls into the class
of the so-called SHTC (symmetric hyperbolic thermodynamically compatible)
systems \cite{GodRom1995,GodRom1996,Rom1998,SHTC-GENERIC}, which are
thermodynamically compatible by construction, i.e. they are over-determined
systems but they automatically posses an additional conservation law which is
the total energy \eqref{energy.cons}. In particular, the structure of the system
\eqref{eqn.PDE} is equivalent to the structure of the system for moving
deformable conductors and dielectrics in \cite{Rom1998,Rom2001} and can be
symmetrized in a similar way using the transformation of variables from
$\bm{q}=\left\{ \MM,\AA,\bm{P},\Burg,\Durg \right\}$ to their thermodynamically
conjugate counterparts $\bm{p} = \left\{ \vv, \bm{\Pi}, \bm{\pi}, \bm{H}, \bm{E}
\right\}$ \eqref{eqn.therm.forces}, and change of the energy potential
$\En(\bm{q})$ to its Legendre transform 
\begin{equation}
	L(\bm{p}) = M_i v^i + \durg{A}{i} E^{A}_{\ i} + \burg{A}{i} H_{iA} + 
	\Pi^k_{\ A} \dist{A}{k} + \pi^A_{\ a} P^a_{\ A} - \En.
\end{equation}

For the symmetric hyperbolicity of the system \eqref{eqn.PDE}, it is sufficient that
the new potential $L(\bm{p})$ is a convex
function of its arguments, which is equivalent to the convexity of the original
energy potential $ \En(\bm{q}) $ due to the properties of
the Legendre transform. Checking the convexity of $ \En $ is not an easy task in
the whole space of state vector, however, for small elastic deformations, the
convexity can be verified by linearizing the energy around the stress-free state
$\dist{A}{k} = \delta^A_{\ k}$, $P^a_{\ A} = \delta^a_{\ A}$, $\durg{A}{i} = 0$,
and $\burg{A}{i} = 0$, and checking the positive definiteness of the Hessian
matrix of the energy potential at this state. It has appeared that the
eigenvalues of the energy Hessian can be computed analytically, and their positiveness is guaranteed provided that the following conditions are satisfied:
\begin{subequations}\label{eqn.convexity.conditions}
	\begin{gather}   
		\rho _0>0, \quad \mu >0, \quad \epsilon >0, \quad c_0>0, \quad 
		c_s>\frac{c_0}{\sqrt{6}}, \\
		\quad C_0>\frac{c_0}{\sqrt{15}}, 
		\quad C_s>\frac{1}{2} \sqrt{\frac{c_0^2}{3}+C_0^2}.
	\end{gather}
\end{subequations}

\subsection{Stress tensor}

A specific form of the stress tensor \eqref{stress} depends on the choice of the
energy potential $ \En $ via the thermodynamic forces \eqref{eqn.therm.forces}.
After the energy potential has been defined, we can now give the explicit
form of the thermodynamic forces. Thus, using the energy potential defined in the
previous section, the thermodynamic forces read
\begin{subequations}\label{eqn.forces.explicit}
	\begin{align}
 E^A_{\ k} = & \frac{1}{\epsilon} \durg{A}{k} + \frac{1}{\rho} \varepsilon_{kij} M^i \burg{A}{j}, 
		\\[2mm]
 H_{kA} = & \frac{1}{\mu} \burg{A}{k} - \frac{1}{\rho} \varepsilon_{kij} M^i \durg{A}{j}, 
		\\[2mm]
 \Pi^k_{\ A} = & \left( \Gamma \En_1 +\En_2 + \En_3 +\En_4 - \En_6 -\En_7 \right) F^k_{\ A} + \rho C_s^2 \dist{A}{i} G'^{ik},
		\\[2mm]
 \pi^A_{\ a} = & \rho \frac{c_0^2}{\gamma} \det(\bm{P})^{\gamma-1} F^{A}_{\ a} + \rho c_s^2 P^{a}_{\ B} g'^{AB},
	\end{align}
\end{subequations}
where $F^k_{\ A}$ and $F^{A}_{\ a}$ are the inverses of the macroscopic
$\dist{A}{k}$ and microscopic $P^a_{\ A}$ distortions respectively, while $
\bm{G}' $ and $ \bm{g}' $ are the deviatoric parts of the
macroscopic and microscopic metric tensors respectively, and $\bm{G}$, and
$\bm{g}$. Also,
$\eta^{ik}$ and $\eta^{AB}$ are the inverses of $\eta_{ik}$ and $\eta_{AB}$
respectively.

It is important to note that the stress tensor $\Sigma^k_{\ i}$ \eqref{stress}
resulting from our definition of the energy is not symmetric. Nevertheless, the
total angular momentum of the system is still conserved because the overall
momentum flux tensor is in fact symmetric, as can be verified by direct
calculation. This is due to the fact that the advection term
\eqref{eqn.PDE.extend.M} is not $\rho v_i v^k$, which is obviously symmetric,
but the non-symmetric tensor $M_i v^k$, which includes contributions from the
torsion fields and keeps the overall momentum flux tensor symmetric.  

\section{Dispersion analysis}\label{sec.DR}

In this section, we analyze the dispersion relation of the proposed model for
small amplitude waves and demonstrate, for certain values of the model
parameters, the presence of a complete frequency band gap.

We shall use the following notations
\begin{equation}
	\bm{q} = \{\MM,\AA,\bm{P},\Burg,\Durg\}
\end{equation}
for the vector of conservative variables, and
\begin{equation}
	\bm{w} = \{\vv,\AA,\bm{P},\Burg,\Durg\}
\end{equation}
for the vector of the so-called primitive variables. With these notations, the
system \eqref{eqn.PDE} can be written in the following quasilinear form
\begin{equation}
	\mathbb{A}\frac{\pd \bm{w}}{\pd t} 
	+ \mathbb{B}^k(\bm{w}) \frac{\pd \bm{w}}{\pd x^k} 
	= \bm{R}(\bm{w}),
\end{equation}
where $\mathbb{A} = \pd \bm{q}/\pd \bm{w}$ is the Jacobian matrix of the
transformation from primitive to conservative variables, and $\bm{R}(\bm{w})$ is
the vector of all the relaxation source terms. The quasilinear form can be
eventually be written in the standard form
\begin{equation}\label{eqn.quasilinear.standard}
	\frac{\pd \bm{w}}{\pd t} 
	+ \mathbb{C}^k(\bm{w}) \frac{\pd \bm{w}}{\pd x^k} 
	= \bm{S}(\bm{w}),
\end{equation}
with $\mathbb{C}^k(\bm{w}) = \mathbb{A}^{-1}\mathbb{B}^k(\bm{w})$ and $\bm{S}(\bm{w}) =
\mathbb{A}^{-1}\bm{R}(\bm{w})$.

To perform the dispersion analysis, we linearize the governing equations
\eqref{eqn.PDE} around a homogeneous equilibrium state at rest $\bm{w}_0$ defined by
constant values of all fields:
\begin{equation}
	\vv = 0, \quad \AA = \bm{I}, \quad \bm{P} = \bm{I}, \quad \Burg = 0, \quad \Durg = 0,  
\end{equation}
where $\bm{I}$ is the identity matrix. Note that at this equilibrium state, all
thermodynamic forces \eqref{eqn.forces.explicit} vanish:
\begin{equation}
	\bm{E}(\bm{w}_0) = 0, \quad \bm{H}(\bm{w}_0) = 0, \quad \bm{\Pi}(\bm{w}_0) = 0, \quad \bm{\pi}(\bm{w}_0) = 0,
\end{equation}
and so does the source term vector, $\bm{S}(\bm{w}_0) = 0$ 

We now restrict the consideration to one-dimensional motions only (however, all
the vector and tensor fields will keep all their three-dimensional components)
and hence, we keep only one spatial coordinate from now on, $x := x^1$,
$\mathbb{C} := \mathbb{C}^1(\bm{w}_0)$. The linearized equations around
$\bm{w}_0$, i.e. $\bm{w}(t,x) = \bm{w}_0 + \tilde{\bm{w}}(t,x)$, can be written
in the form
\begin{equation}
	\frac{\partial \tilde{\bm{w}}}{\partial t} 
	+ \mathbb{C} \frac{\partial \tilde{\bm{w}}}{\partial x} 
	= \mathbb{S}\tilde{\bm{w}},
\end{equation}
where $\mathbb{S} = \mathbb{A}^{-1}\pd \bm{R}/\pd \bm{w}$ is the Jacobian matrix
of the source vector, evaluated at the equilibrium state $\bm{w}_0$.

\subsection{Plane wave solution}

In what follows, we shall drop the tilde sign in $\tilde{\bm{w}}$ and write it
just as $\bm{w}$. Then, we look for plane wave solutions of the linearized
equations in the form
\begin{equation}
	\bm{w} = \hat{\bm{w}} e^{i(k x - \omega t)},
\end{equation}
where $\hat{\bm{w}}$ is the constant amplitude vector, $k$ is the wave number
(assumed to be real), $\omega$ is the angular frequency (might be complex in
general), and $i = \sqrt{-1}$ is the imaginary unit. Substituting this ansatz
into the linearized equations, we obtain an eigenvalue problem 
\begin{equation}
	\left( k \mathbb{C} + i \mathbb{S} - \omega \mathbb{I} \right) \hat{\bm{w}} = 0,
\end{equation}
or
\begin{equation}
	\left( \mathbb{C} + \frac{i}{k} \mathbb{S} - \lambda \mathbb{I} \right) \hat{\bm{w}} = 0,
\end{equation}
which has nontrivial solutions only if
\begin{equation}\label{eqn.dispersion.relation}
	p(k,\lambda) := \det \left( \mathbb{C} + \frac{i}{k} \mathbb{S} 
	- \lambda \mathbb{I} \right) = 0,
\end{equation}
where $\lambda(k) = \omega(k)/k$ is the phase velocity.

As we discussed in Sec.\,\ref{sec.energy.cons}, there are no dissipative
processes in the model, and all the source terms are of reversible nature.
Therefore, all eigenvalues $\lambda(k)$ must be real. 

It has appeared that the polynomial $p(k,\lambda)$ can be factorized into four
polynomials (we ignore the zero eigenvalue associated with the non-propagating
modes which has multiplicity of 15):
\begin{equation}\label{eqn.p}
	p(k,\lambda) = p_1(k,\lambda) \, p_2(k,\lambda) \, p_3(k,\lambda) \, 
	p_4(k,\lambda),
\end{equation}
where all polynomials $p_j(k,\lambda)$ have only even powers of $\lambda$.
Moreover, $p_1(k,\lambda)$ and $p_2(k,\lambda)$ are polynomials of degree two,
and $p_3(k,\lambda)$ and $p_4(k,\lambda)$ are polynomials of degree six in
$\lambda$. This means that in principle, the roots can be found analytically.
However, their expression is quite lengthy.

Also, note that for high wave numbers $k \to \infty$, the eigenvalue problem
\eqref{eqn.dispersion.relation} reduces to the eigenvalue problem for matrix
$\mathbb{C}$ with $\lambda$ having the following values:
\begin{equation}\label{eqn.c}
	\pm C_l, 
	\qquad 
	\pm C_s, 
	\qquad 
	\pm \cinf = \pm \frac{1}{\sqrt{\epsilon\mu}},
\end{equation}
where $C_l = \sqrt{C_0^2 + \frac{4}{3} C_s^2}$ and $C_s$ are the speeds of the
longitudinal and transverse acoustic waves in the bulk material, while $\cinf$
corresponds to the speed of propagation of rotational modes associated with
relative rotations of the microscopic and macroscopic frames. 

Overall, we will distinguish three types of waves in the medium: the
\textit{rotational} modes (roots of $p_1(k,\lambda)$ and $p_2(k,\lambda)$), the
\textit{longitudinal} modes (roots of $p_3(k,\lambda)$), and the \textit{shear}
modes (roots of $p_4(k,\lambda)$). Additionally, we will refer to the modes that
have a non-zero frequency at $k \to 0$ as \textit{optical} modes, while the
modes that have zero frequency at $k \to 0$ will be called \textit{acoustic}
modes. As we shall see later, all branches of the rotational modes are always
optical, while the longitudinal and shear modes have branches of both types,
one acoustic branch and two optical branches.

The four roots (2 positive and 2 negative) of the polynomials $p_1(k,\lambda)$
and $p_2(k,\lambda)$ in \eqref{eqn.p} read
\begin{subequations}\label{eqn.roots.p1p2}
	\begin{align}
		\lambda(k)^2 &= c^2+\frac{\rho _0 \left(c_0^2 (2 \alpha +\beta )+3 \beta  C_0^2\right)}{6 \alpha ^2 \beta \epsilon k^2 }, \\
		\lambda(k)^2 &= c^2+\frac{\rho _0 \left(4 \alpha ^2 c_s^2-\frac{1}{3} \beta  c_0^2 (2 \alpha +\beta )-\beta ^2 \left(C_0^2-4 C_s^2\right)\right)}{2 \alpha ^2 \beta ^2 \epsilon k^2 },
	\end{align}
\end{subequations}
or if we use that $\alpha = \beta$,
\begin{subequations}\label{eqn.roots.p1p2.2}
	\begin{align}
		\lambda(k)^2 &= c^2+\frac{\rho _0 \left(c_0^2+C_0^2\right)}{2 \alpha ^2 \epsilon k^2}, \\
		\lambda(k)^2 &= c^2-\frac{\rho _0 \left(c_0^2+C_0^2-4 \left(c_s^2+C_s^2\right)\right)}{2 \alpha ^2 \epsilon k^2 },
	\end{align}
\end{subequations}
The roots of the polynomials $p_3(k,\lambda)$ and $p_4(k,\lambda)$ are quite
lengthy and will not be reported here.

\subsection{Cut-off frequencies}
	
We introduce the cutoff frequencies, i.e. the frequencies at which
the wave number $k$ vanishes:
\begin{subequations}\label{eqn.cutoff.freqs}
	\begin{align}
		\omega_\infty & = \sqrt{\frac{\rho_0 \left(c_0^2 (2 \alpha +\beta ) + 
		4 \alpha  c_s^2\right)}{3 \alpha  \beta ^2 \epsilon }},\\
		\omega_0 & = \sqrt{\frac{\rho_0 \left(c_0^2 (2 \alpha +\beta ) + 3 \beta  C_0^2\right)}{6 \alpha ^2 \beta  \epsilon }}, \\
		\omega_s  & =
		\sqrt{\frac{\rho_0 \left(12 \alpha ^2 c_s^2-\beta  \left(c_0^2 (2 \alpha +\beta )+3 \beta  \left(C_0^2-4 C_s^2\right)\right)\right)}{6 \alpha ^2 \beta ^2 \epsilon }} ,\\
		\omega_l  & = \sqrt{\frac{\rho_0 \left(\frac{1}{3} c_0^2 (2 \alpha +\beta ) (3 \alpha +\beta )+4 \beta ^2 C_0^2\right)}{\alpha ^2 \beta ^2 \epsilon }},
	\end{align}
\end{subequations}
or, if we use that $\alpha = \beta $, we can rewrite them as 
\begin{subequations}\label{eqn.cutoff.freqs2}
	\begin{align}
		\omega_\infty & = \sqrt{\frac{\rho_0 \left(c_0^2+\dfrac{4 c_s^2}{3}\right)}{\alpha ^2 \epsilon }} ,\\
		\omega_0 & = \sqrt{\frac{\rho_0 \left(c_0^2+C_0^2\right)}{2 \alpha ^2 \epsilon }}, \\
		\omega_s & = \sqrt{-\frac{\rho_0 \left(c_0^2+C_0^2-4 \left(c_s^2+C_s^2\right)\right)}{2 \alpha ^2 \epsilon }},\\
		\omega_l & = 2 \sqrt{\frac{\rho_0 \left(c_0^2+C_0^2\right)}{\alpha ^2 \epsilon }}.
	\end{align}
\end{subequations}
As we shall see later, $\omega_0$, $\omega_l$ and $\omega_s$ are the frequencies
that correspond to the cutoff frequencies ($k=0$) of the optical, longitudinal
and shear modes, while $\omega_\infty$ is the frequency of the acoustic
longitudinal mode at $k \to \infty$. Our numerical experiments show that the
width of the frequency band gap is defined as $\omega \in [\omega_\infty,
\omega_0]$, see Fig.\,\ref{fig.w_vs_k}.

We also introduce the following equilibrium (i.e. when $\omega \to 0$) longitudinal and transverse
velocities:
\begin{subequations}\label{eqn.Vl_Vs}
	\begin{align}
		V_l^2 & = \frac{3 c_0^2 C_0^2 \alpha (2 \alpha +\beta )}{c_0^2 (2 \alpha +\beta ) (3 \alpha +\beta )+12 \beta ^2 C_0^2} + \frac{4}{3} V_s^2,\\
		V_s^2 & = \frac{4 c_s^2 C_s^2 \alpha^2}{4 c_s^2 \alpha^2 
		- \left(C_0^2-4 C_s^2\right)\beta^2
		- \dfrac{1}{3} c_0^2 (2 \alpha +\beta ) \beta },\\
	\end{align}
\end{subequations}
or, if $\alpha = \beta$,
\begin{subequations}\label{eqn.Vl_Vs.2}
	\begin{align}
		V_l^2 & = \frac{3 c_0^2 C_0^2}{4 (c_0^2 + C_0^2)} + \frac{4}{3} V_s^2,\\
		V_s^2 & = \frac{4 c_s^2 C_s^2}{- c_0^2 - C_0^2 + 4 \left( c_s^2 + C_s^2\right)}.
	\end{align}
\end{subequations}
These are the phase velocities of the acoustic modes at low frequencies $\omega \to 0$.

\subsection{Numerical results and discussion}\label{sec.results}

In this section, we plot the dispersion curves of the proposed model for a
certain set of material parameters given in Table\,\ref{tab.model_parameters}.
The values of these parameters are chosen similar to those used in
\cite{Madeo2015a}, where possible. The main trend is that the microstructure is
softer (lower elastic modulus, or sound speeds in our case) than the macroscopic
material. The microstructure characteristic length scale $\ell$ is chosen to be
equal to $0.002$\,m, as in \cite{Madeo2015a}.

Despite the fact that the roots of the polynomials in \eqref{eqn.p} can be found
analytically, their expressions (for $p_3(\lambda,k)$ and $p_4(\lambda,k)$) are
quite lengthy and due to the accumulation of round-off errors, spurious
imaginary parts of the roots may appear. Instead, we compute the roots
numerically using the \texttt{roots} function from the \texttt{Matlab} software
\cite{MATLAB} which always results in real numerical roots if the analytical
roots are real.

\begin{table}[h!]
    \centering
    \begin{tabular}{|c|c|c|}
    \hline
    {Parameter} & {Value} & {Physical Unit} \\
    \hline
    $\rho_0$ & $2000$ & $\text{kg}/\text{m}^3$ \\
    $C_0$ & $600$ & $\text{m}/\text{s}$ \\
    $C_s$ & $600$ & $\text{m}/\text{s}$ \\
	$C_l$ & $916.5$ & $\text{m}/\text{s}$ \\
    $c_0$ & $100$ & $\text{m}/\text{s}$ \\
    $c_s$ & $100$ & $\text{m}/\text{s}$ \\
    $\epsilon$ & $2 \times 10^{-5}$ & $\text{kg} \cdot \text{m}$ \\
    $\mu$ & $5 \times 10^{-1}$ & $\text{s}^2/(\text{m}^3 \cdot \text{kg})$ \\
	$\cinf = 1/\sqrt{\epsilon\mu}$ & $316.23$ & $\text{m}/\text{s}$ \\
	$\ell$ & $ 2\cdot10^{-3}$ & $\text{m}$ \\
    $\alpha=1/\ell$ & $100$ & $1/\text{m}$ \\
    $\beta=1/\ell$ & $100$ & $1/\text{m}$ \\
    \hline
    \end{tabular}
    \caption{Model parameters, their values and physical units.}
    \label{tab.model_parameters}
\end{table}

\begin{figure}[hbt!]
	\centering
	\includegraphics[width=0.90\textwidth]{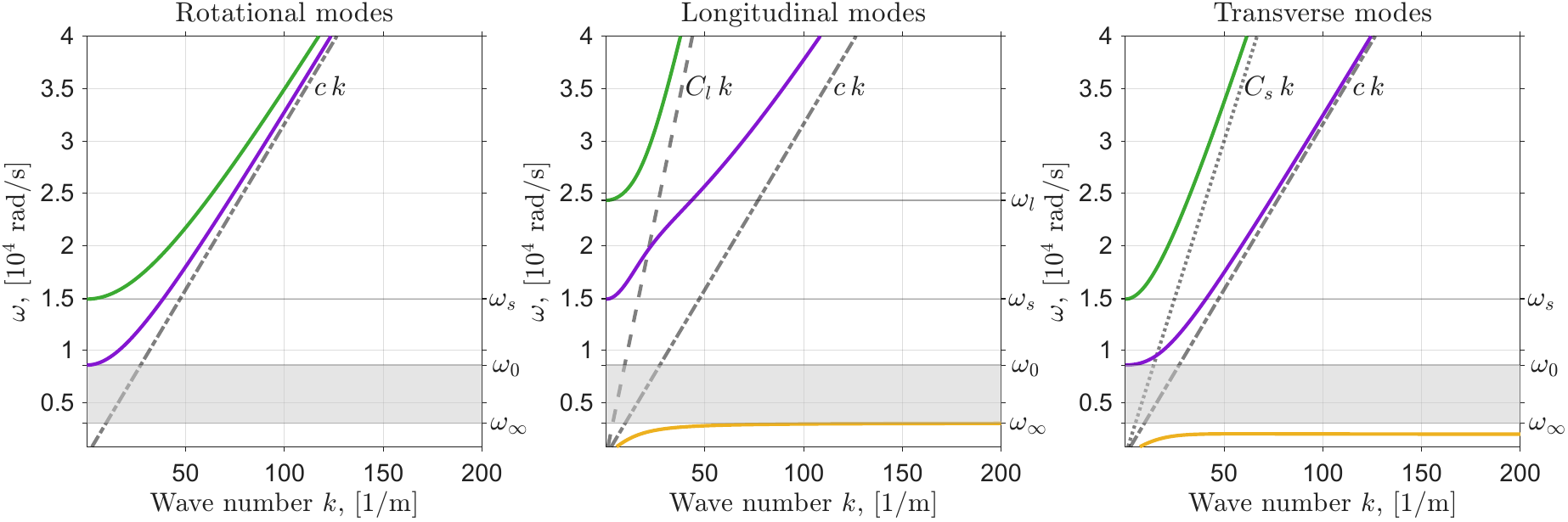}
	\caption{Dispersion curves $\omega(k)$ of the proposed model for
		microstructured solids. The parameters used to generate these curves are
		given in Table\,\ref{tab.model_parameters}. The complete frequency band
		gap is shown as the gray shaded rectangle. The acoustic modes are shown in
		yellow, while the optical modes are shown in green (fast) and purple
		(slow).}
	\label{fig.w_vs_k}
\end{figure}

Fig.\,\ref{fig.w_vs_k} shows the dispersion curves $\omega(k) = k \lambda(k)$
(recall that all eigenvalues $\lambda(k)$ are real) for the parameters given in
Table\,\ref{tab.model_parameters}. In the left figure, we plot positive roots of
the first two polynomials $p_1$ and $p_2$ in \eqref{eqn.p}, which correspond to
the rotational modes. In the middle and right figures, we plot positive roots of
the polynomials $p_3$ and $p_4$, corresponding to the longitudinal and shear
modes respectively. We can see that all branches of the rotational modes are
optical, while the longitudinal and shear modes have branches of both types, one
acoustic branch (in yellow) and two optical branches (fast, in green, and slow,
in purple). A frequency band gap is clearly visible (gray shaded rectangle) in
all figures between the asymptotic frequency $\omega_\infty$ of the acoustic
longitudinal mode and the cutoff frequency of the optical rotational mode
$\omega_0$. The width of the band gap is approximately equal to $\omega_0 -
\omega_\infty \approx 2.8\times 10^4$\,rad/s.

\begin{figure}[hbt!]
	\centering
	\includegraphics[width=0.90\textwidth]{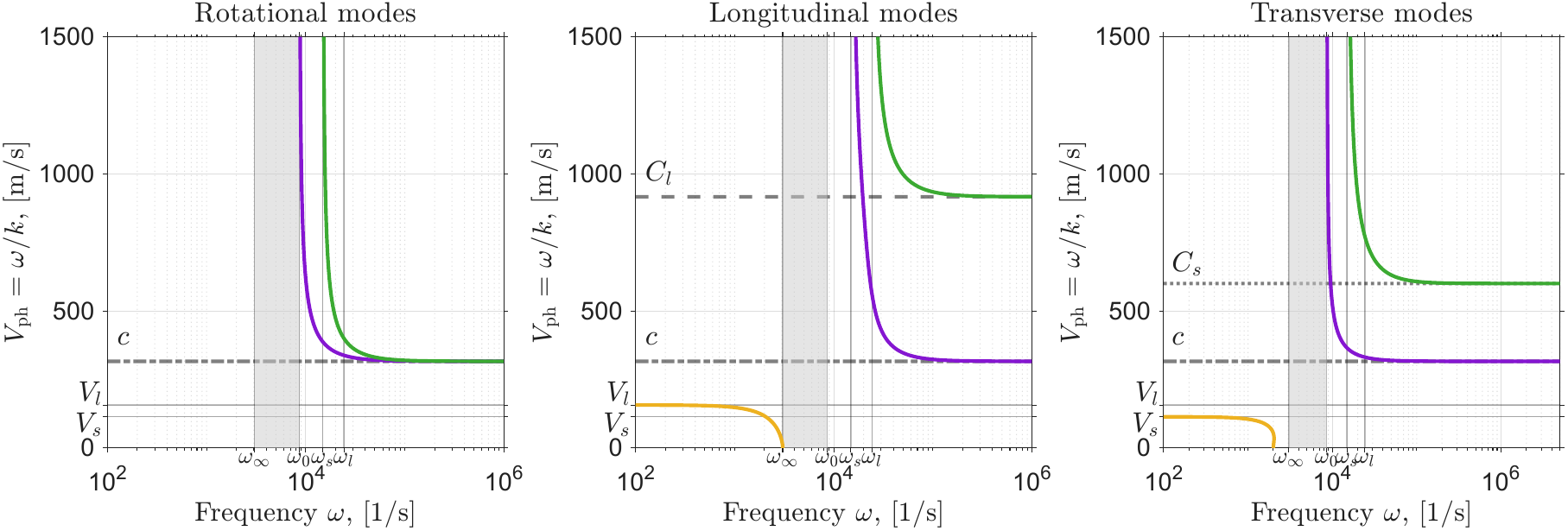}
	\caption{Phase velocities $V_{\text{ph}}(\omega) = \omega/k$ of the proposed
		model and the material parameters from
		Table\,\ref{tab.model_parameters}. The complete frequency band gap is
		shown as the gray shaded rectangle. The acoustic modes are shown in
		yellow, while the optical modes are shown in green (fast) and purple
		(slow).}
	\label{fig.Vph_vs_w}
\end{figure}

Fig.\,\ref{fig.Vph_vs_w} shows the same dispersion curves as in
Fig.\,\ref{fig.w_vs_k}, but plotted as phase velocity $V_{\text{ph}}(\omega) =
\lambda$ as a function of frequency $\omega = k \lambda$ for the same set of
material parameters. The frequency band gap is again clearly visible and shown
as the gray shaded rectangle. One can see that, in the low frequency domain
($\omega \to 0 $), there are single longitudinal and transverse waves with phase
velocities $V_l$ and $V_s$ (yellow lines), as given by equations
\eqref{eqn.Vl_Vs} and \eqref{eqn.Vl_Vs.2}. This means that the material behaves
as a simple (no microstructure) elastic solid in the low frequency regime.
However, its sound speeds, $V_l$ and $V_s$, are different from the macroscopic
$C_l = \sqrt{C_0^2 + 4/3 C_s^2}$ and $C_s$ and microscopic $c_l = \sqrt{c_0^2 +
4/3 c_s^2}$ and $c_s$ material constants. 

Moving in the high frequency domain (or short wavelengths), we first encounter
the frequency band gap where no wave propagation is possible. This simply means
that the waves with frequencies in the band gap $\omega \in [\omega_\infty,
\omega_0]$ are evanescent waves (or standing waves) which oscillate in a
localized region of space.

After the band gap, $\omega > \omega_0$, we can see that there are fast (green)
and slow (purple) waves for every mode (rotational, longitudinal and shear).
However, in the interval $\omega \in [\omega_0, \omega_s]$, there are only slow
rotational and shear waves, while slow longitudinal waves appear only for
$\omega > \omega_s$. Finally, for high frequencies $\omega \to \infty$, all
phase velocities tend to the characteristic velocities of the homogeneous
system given in \eqref{eqn.c}.

\begin{figure}[hbt!]
	\centering
	\includegraphics[width=0.90\textwidth]{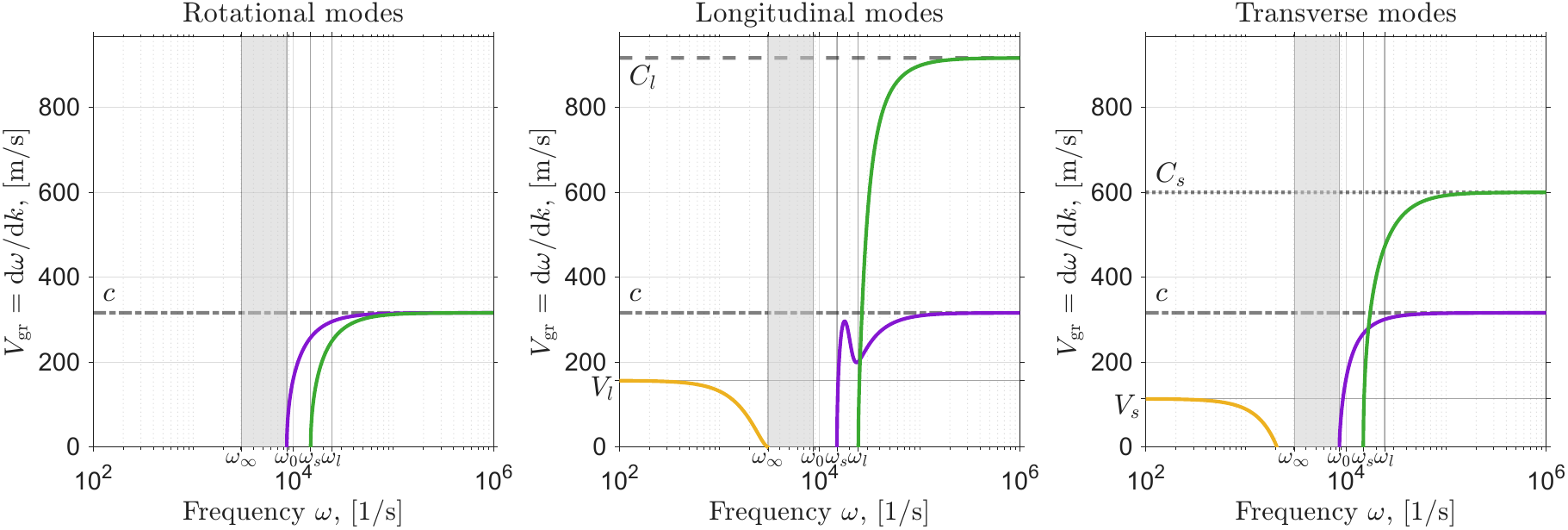}
	\caption{Group velocities $V_{\text{gr}}(\omega) =
		\mathrm{d}\omega/\mathrm{d}k$ of the proposed model and the material
		parameters from Table\,\ref{tab.model_parameters}. The complete
		frequency band gap is shown as the gray shaded rectangle. The acoustic
		modes are shown in yellow, while the optical modes are shown in green
		(fast) and purple (slow).}
	\label{fig.Vgr_vs_w}
\end{figure}

Note that the studied model is a system of hyperbolic equations which is a
standard tool for modeling wave propagation phenomena. That means that the
system must admit only finite propagation velocities (causality). In this
context, a remarkable feature of the phase velocity curves $V_\text{ph}(\omega)$
in Fig.\,\ref{fig.Vph_vs_w} is the unlimited growth of the phase velocity at the
cutoff frequencies $\omega_0$ (band gap edge), $\omega_s$, and $\omega_l$. This
is of course a direct consequence of the fact that at these frequencies, the
wave number $k$ vanishes, $k \to 0$, while the frequency $\omega$ remains
finite, leading to $V_{\text{ph}} = \omega/k \to \infty$. This however does not
contradict the principle of causality because the information and energy is
transported at the group velocities $V_{\text{gr}} =
\mathrm{d}\omega/\mathrm{d}k$ which, for the studied model, remain finite and
bounded as can be seen in Fig.\,\ref{fig.Vgr_vs_w}. The curves on this figure
are obtained by applying central finite differences to the dispersion curves
$\omega(k)$ from Fig.\,\ref{fig.w_vs_k}.

\subsection{Influence of the parameters $\alpha$ and $\beta$ on the band gap}

Unlike conventional dispersive models which are based on higher-order time and
space derivatives, the proposed model is based on first-order hyperbolic PDEs
with algebraic relaxation source terms. In fact, in such a framework, the
dispersive properties of a model are encoded in the source terms which have a
special antisymmetric structure, see \eqref{eqn.sum.source}. For example,
first-order hyperbolic equations were applied to model dispersive phenomena in
\cite{Gavrilyuk2005,Romenski2011,Mazaheri2016,Favrie2017a,Lombard2019,Dhaouadi2019,
Dhaouadi2022a,SHTC_surfacetension2025}, and in particular, for modeling the
frequency band gap in acoustic metamaterials \cite{Lombard2019}. It is,
therefore, interesting to study the influence of the relaxation parameters
$\alpha$ and $\beta$ (which appear only in the source terms) on the dispersion
curves and on the band gap.

It is clear that if one set $\alpha = \infty$ and $\beta = \infty$, the source
terms vanish and the model reduces to a non-dispersive model with wave
propagating at constant speeds given in \eqref{eqn.c}.

As we have seen, the width of the complete band gap is defined as $\omega \in
[\omega_\infty, \omega_0]$. Hence, to close the band gap, one needs to make
$\omega_\infty$ and $\omega_0$ equal. Thus, by lowering $\beta$ from
$\beta=\alpha$ to 
\begin{equation}
	\beta_\mathrm{crit} = 2 \alpha  \sqrt{\frac{c_0^2+2 c_s^2}{c_0^2+3 C_0^2}}
\end{equation}
one can gradually decrease the band gap width until it disappears at $\beta =
\beta_\mathrm{crit}$.

On the other hand, by increasing $\beta$ to infinity, the cutoff frequency
$\omega_\infty$ tends to zero, $\omega_\infty \to 0$, which also results in the
vanishing of the acoustic longitudinal and transverse modes (yellow curves
in Fig.\,\ref{fig.w_vs_k}--\ref{fig.Vgr_vs_w}).

\begin{figure}[hbt!]
	\centering
	\includegraphics[width=0.90\textwidth]{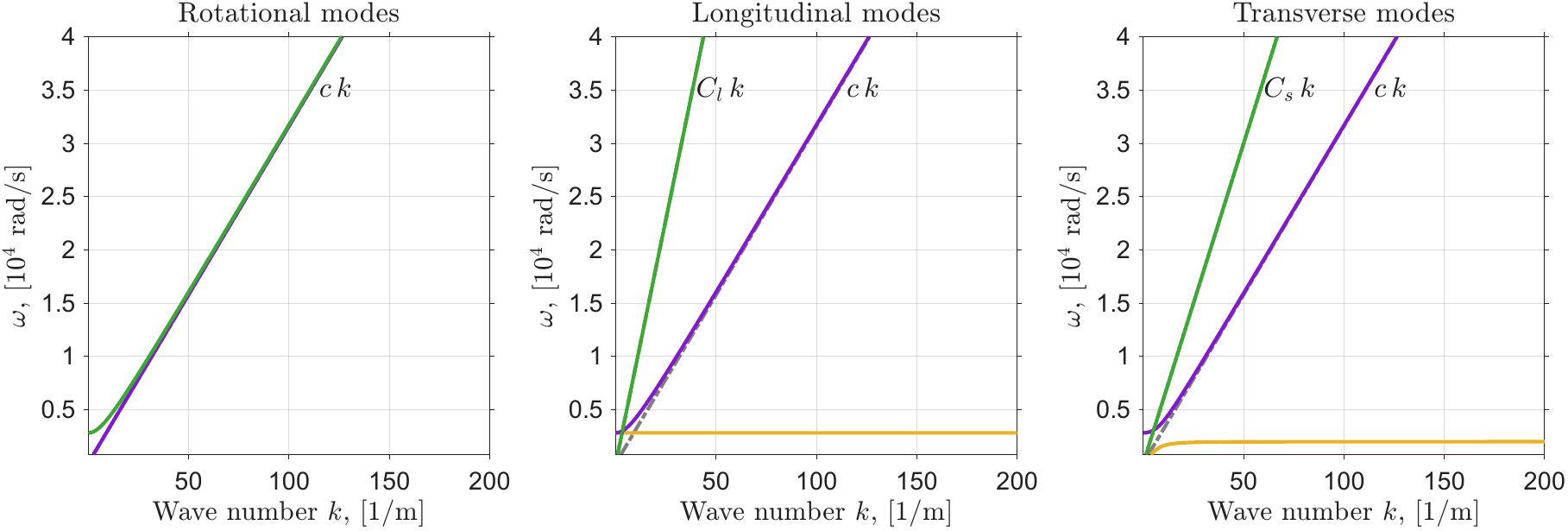}
	\caption{Dispersion curves $\omega(k)$ of the proposed model for the case
	$\alpha = \infty$ (the relaxation terms $\frac{1}{\alpha} \Pi^k_{\ A}$ and
	$\frac{1}{\alpha} E^A_{\ k}$ are switched off) and other parameters as in
	Table\,\ref{tab.model_parameters}. The band gap is absent, i.e. for every
	$\omega$ there is a wave number $k$.}
	\label{fig.w_vs_k_alpha_inf}
\end{figure}

Finally, setting $\alpha = \infty$, i.e. switching off the relaxation terms
multiplied by $1/\alpha$, while keeping $\beta$ finite, results in the
dispersion curves shown in Fig.\,\ref{fig.w_vs_k_alpha_inf}. One can see that
for every $\omega$ there is a wave number $k$, i.e. in this case, the band gap
disappears as well.

\subsection{Comparison with Madeo et al work}\label{sec.comparison.Madeo}

This work is motivated by the model for microstructured solids developed by
Madeo et al \cite{Madeo2015a}, where it was demonstrated that their model can
exhibit a complete frequency band gap. In this section, we discuss similarities
and differences between the model presented here and the model by Madeo et al
\cite{Madeo2015a}. 

Despite the fact that the model in \cite{Madeo2015a} is formulated for the case
of small elastic deformations while our model is formulated for finite strains,
both models share some key similarities. First of all, both models are
variational models and therefore one can compare them at the level of the
Lagrangian density, or the energy potential. Let us recall the energy potential
from \cite{Madeo2015a} in which we omit the elastic energies of the macro and
micro distortions for the sake of brevity (the elastic energies, our
\eqref{eqn.elast.macro} and \eqref{eqn.elast.micro}, are similar in both models
and do not play a key role in the emergence of the band gap). The energy
potential from \cite{Madeo2015a} reads
\begin{equation}\label{eqn.En.Madeo}
	\En = \frac{\rho}{2} \Vert \vv \Vert^2 
	+ \frac{\eta}{2} \Vert \dot{\bm{P}} \Vert^2
	+ \mu_c \Vert \mathrm{skew}\left( \nabla\bm{u} - \bm{P} \right) \Vert^2
	+ \frac{\alpha_c}{2} \Vert \nabla \times \bm{P} \Vert^2,
\end{equation}
where the dot denotes the material time derivative, ``skew'' stands for the
skew-symmetric part of a tensor, $\eta$, $\mu_c$ and $\alpha_c$ are positive
model parameters, $\bm{u}$ is the total displacement field, while $\bm{P}$
stands for the microdistortion field. For small deformations one can assume that
$\Dist = \nabla \bm{u} - \bm{P}$ is the macrodistortion.

Despite the use of the space-time gradients of the microdistortion $\bm{P}$ in
\cite{Madeo2015a} and our use of the space-time gradients of the
macrodistortion, the role of the terms $\frac{\eta}{2} \Vert \dot{\bm{P}}
\Vert^2$ and $\frac{\alpha_c}{2} \Vert \nabla \times \bm{P} \Vert^2$ in
\eqref{eqn.En.Madeo} is similar to the role of our terms $\frac{1}{2\epsilon}
\Vert \Durg \Vert^2$ and $\frac{1}{2\mu} \Vert \Burg \Vert^2$ in \eqref{eqn.En},
i.e. to represent the microinertia effect and to account for the geometric
incompatibilities in the material.

More importantly, one of the key findings in \cite{Madeo2015a} was the role of
the so-called Cosserat couple modulus, $\mu_c$ in \eqref{eqn.En.Madeo}. Thus,
according to \cite{Madeo2015a}, the presence of a complete frequency band gap in
their model is possible only if $\mu_c > 0$. On the other hand, in our model
similar role is played by the relaxation parameters $\alpha$ and $\beta$. To
better understand similarities between these key parameters of the two models,
let us note that in \cite{Madeo2015a}, the governing equation for the
microdistortion $\bm{P}$ is a second-order in time PDE which can be seen as a
first order PDE on $\dot{\bm{P}}$. Therefore, its role is similar to our PDE
\eqref{eqn.PDE.extend.D} for the $\Durg$ field. Moreover, in \cite{Madeo2015a},
there is a source term $\sim \mu_c \mathrm{skew}\left( \nabla\bm{u} - \bm{P}
\right) $ in the PDE for $\bm{P}$, which triggers time variations $\dot{\bm{P}}$
as soon as the orientations of the frames $\nabla\bm{u}$ and $\bm{P}$ are not
aligned. Likewise, disorientation between the macro $\Dist$ and micro $\bm{P}$
frames (encoded in the associated macro $\Pi^k_{\ A}$ and micro $\pi^A_{\ a}$
stress tensors) governs the evolution of the $\Durg$ field in our model, and
subsequently controls the time variation of the $\Dist$ field. Therefore, the
Cosserat couple modulus $\mu_c$ in \cite{Madeo2015a} plays a very similar role
to our relaxation parameters $\alpha$ and $\beta$.

At the end of this section, we emphasize important differences between the two
models. First of all, our model is formulated for finite deformations while the
model in \cite{Madeo2015a} is valid only for small deformations. This makes our
model more general and applicable to a wider range of problems, including the
study of nonlinear waves in microstructured solids. Secondly, we would like to
stress the clear geometrical meaning of our model within the Riemann-Cartan
geometry with non-vanishing torsion. This geometrical framework allows for a
more systematic derivation of extensions of the presented model via including
the curvature and non-metricity tensors. Finally, we emphasize the similarities
between our model for microstructured solids and the Maxwell equations of
electromagnetism on the level of energy potential and the structure of the
governing equations (in particular the equations for $\Durg$ and $\Burg$ fields,
and the momentum equation). This analogy might be useful to link the wave
propagation phenomena in photonic and phononic metamaterials.

\section{Conclusion and outlook} \label{sec.conclusion}

We have presented a new model for elastic microstructured solids that can
undergo finite deformations. However, in this paper, we have restricted our
attention to the case of small deformations only to perform the dispersion
analysis and demonstrate the presence of a complete frequency band gap in the
dispersion relation. 

The important properties of the presented model are that it is formulated within
the Riemann-Cartan geometry with non-vanishing torsion, it is hyperbolic and
fully compatible with the first law of thermodynamics. Moreover, the
governing equations are symmetrizable, i.e. they can be written as a symmetric
hyperbolic system, subjected to the total energy of the system is a convex
potential. In particular, the energy potential used in this paper is convex at
least in a vicinity of the equilibrium state.

The material was considered to have two time and length scales: the macroscopic
scale associated with the observation scale (limited by our resolution
capabilities), and the microscopic scale associated with the material
microstructure, which is assumed to be unavailable for direct observations. The
microstructure was not specified in this work, but it can be for example a
lattice structure in acoustic metamaterials, a network of cracks, or a system of
dislocations.

The state of the material was defined by two frame fields: the macroscopic
distortion field which describes the deformation and rotation of the macroscopic
material element, and the microdistortion field which describes the deformation
and rotation of the microstructure elements. The coupling of the two scales
occurs via the assumption that macroscopic distortion field is not geometrically
compatible due to independent deformation of the microstructure. This
results, even in the case of pure elastic deformations, in the presence of
instantaneous geometric incompatibilities in the material which are modeled via
the four-dimensional torsion tensor associated with the macroscopic frame
field.

The dispersion analysis performed in Section\,\ref{sec.DR} for
small amplitude waves has shown that the proposed model can exhibit a complete
frequency band gap for certain values of the model parameters. The band gap is a
direct consequence of the presence of the microstructure and the two-scale
nature of the material. The width of the band gap can be controlled by the
relaxation parameters $\alpha$ and $\beta$ which govern the rate of energy
exchange between the macroscopic and microscopic scales.

In Section\,\ref{sec.comparison.Madeo}, we have compared our model with the
model by Madeo et al \cite{Madeo2015a} and have shown that despite the different
formulations (finite vs small deformations), both models share some key
similarities. In particular, the role of the Cosserat couple modulus in
\cite{Madeo2015a} is similar to the role of our relaxation parameters $\alpha$
and $\beta$ in controlling the emergence of the frequency band gap. Important
differences between the two models were also discussed in
Section\,\ref{sec.comparison.Madeo}.

Future work will focus on exploration of the analogies between the presented
model and Maxwell's equations of electromagnetism in dispersive media. This
analogy might be useful to link the wave propagation phenomena in photonic and
phononic metamaterials. For example, it is known that electric permittivity and
magnetic permeability of real materials might be second-order tensors. Likewise,
our parameters $\epsilon$ and $\mu$ could be extended to be second-order tensors
as well and linked to the microstructured geometry of the material.
Reconstruction of these parameters, as well as of the relaxation parameters
$\alpha$ and $\beta	$, from experimental data will also be studied in future
works.

We will also focus on the numerical simulation of wave propagation phenomena in
microstructured solids under small and finite deformations using high-order
accurate shock-capturing numerical methods for hyperbolic PDEs.

Finally, note that the model does not explore the full capabilities of the
Riemann-Cartan geometry for microstructured solids
\cite{NguyenLeMarrec2022,CRESPO2025a}. In particular, the four-dimensional
Riemann curvature tensor is assumed to vanish in this work. The curvature tensor
is proportional to the second space-time gradients of the frame field and
therefore it can be used to account for higher degree of heterogeneity of the
material microstructure, e.g. see \cite{Neff2023}. In future work, the inclusion
of the four-dimensional Riemann curvature tensor in the model and its influence
on the wave propagation phenomena will be studied.

\section*{Acknowledgements}
The work of I.P. was partially supported by the European Union Next Generation EU,
Mission 4 Component 2--CUP E53D23005840006, the Italian Ministry of University
and Research (MUR) with the PRIN Project 2022 No. 2022N9BM3N, European Union
under the Italian National Recovery and Resilience Plan (NRRP) of
NextGenerationEU, partnership on "Telecommunications of the Future" (PE00000001
- program "RESTART") (CUP: E63C22002040007).

\printbibliography

\end{document}

%% file: structure2025.tex
\usepackage[hmarginratio=2:1,top=32mm,left=25mm,columnsep=0.5cm]{geometry}

\usepackage{amsmath,amssymb,amsfonts,latexsym,float,graphics,epsfig,amsthm}
\usepackage[utf8]{inputenc} 
\usepackage{url}            
\usepackage{booktabs}       
\usepackage{amsfonts}       
\usepackage{nicefrac}       
\usepackage{microtype}      
\usepackage{graphicx}
\usepackage{doi}
\usepackage[all]{xy}
\usepackage{setspace}
\usepackage[dvipsnames]{xcolor}
\usepackage{fourier}
\usepackage{bm}		
\usepackage{relsize} 
\usepackage{accents}

\usepackage{tikz}
\usetikzlibrary{cd}
\usepackage{hyperref}
\hypersetup{
	 colorlinks=true,             
	 linkcolor=blue,
	 citecolor=NavyBlue,
	 urlcolor=blue }
\usepackage{authblk}  

\usepackage[colorinlistoftodos]{todonotes}

\usepackage[backend=biber,giveninits=true,sorting=nyt,url=false,doi=true,eprint=false,isbn=false,
backref,backrefstyle=none,maxbibnames=99]{biblatex}
\DefineBibliographyStrings{english}{%
	backrefpage = {Cited on p\adddot},%
	backrefpages = {Cited on pp\adddot}%
}
\bibliography{references}

\renewbibmacro{in:}{%
	\ifboolexpr{%
		test {\ifentrytype{article}}%
	}{}{\printtext{\bibstring{in}\intitlepunct}}%
}
\renewcommand{\AA}{\bm{A}}

\newcommand{\MM}{\bm{M}}

\newcommand{\vv}{\bm{v}}

\newcommand{\En}{\mathcal{E}}						%

\newcommand{\kronecker}[2]{\delta^{#1}_{\ #2}}
\newcommand{\durg}[2]{ D_{\ #1}^{#2} }	
\newcommand{\LeviCivitaUp}[1]{\varepsilon^{#1}}

\newcommand{\pd}{\partial}

\newcommand{\dist}[2]{ A^{#1}_{\phantom{#1}#2} }	
\newcommand{\Dist}{ \bm{A} }	
\newcommand{\Plast}{ \bm{P} }	
\newcommand{\Burg}{ \bm{B} }	
\newcommand{\Durg}{ \bm{D} }	

\newcommand{\burg}[2]{ B^{{#1}{#2}} }	

\newcommand{\vel}[1]{v^{#1}}
\newcommand{\Tors}[2]{T^{#1}_{\phantom{a}#2}}

\newcommand{\transpose}{{\mathrm {\mathsmaller T}}}

\newcommand{\HT}[1]{\accentset{\star}{T}^{#1}}

\newcommand{\cinf}{c}	



%% file: references.bib
@article{Neff2023,
  title = {Modeling a Labyrinthine Acoustic Metamaterial through an Inertia-Augmented Relaxed Micromorphic Approach},
  author = {Voss, Jendrik and Rizzi, Gianluca and Neff, Patrizio and Madeo, Angela},
  date = {2023-10},
  journaltitle = {Mathematics and Mechanics of Solids},
  shortjournal = {Mathematics and Mechanics of Solids},
  volume = {28},
  number = {10},
  pages = {2177--2201},
  issn = {1081-2865, 1741-3028},
  doi = {10.1177/10812865221137286},
  url = {https://journals.sagepub.com/doi/10.1177/10812865221137286},
}

@article{DellIsola2017,
  title = {Higher-Gradient Continua: {{The}} Legacy of {{Piola}}, {{Mindlin}}, {{Sedov}} and {{Toupin}} and Some Future Research Perspectives},
  shorttitle = {Higher-Gradient Continua},
  author = {family=Isola, given=Francesco, prefix=dell’, useprefix=true and Corte, Alessandro Della and Giorgio, Ivan},
  date = {2017-04},
  journaltitle = {Mathematics and Mechanics of Solids},
  shortjournal = {Mathematics and Mechanics of Solids},
  volume = {22},
  number = {4},
  pages = {852--872},
  issn = {1081-2865, 1741-3028},
  doi = {10.1177/1081286515616034},
  url = {https://journals.sagepub.com/doi/10.1177/1081286515616034},
}

@article{DellIsola2022,
  title = {Second-Gradient Continua: {{From Lagrangian}} to {{Eulerian}} and Back},
  shorttitle = {Second-Gradient Continua},
  author = {family=Isola, given=Francesco, prefix=dell’, useprefix=true and Eugster, Simon R and Fedele, Roberto and Seppecher, Pierre},
  date = {2022-12},
  journaltitle = {Mathematics and Mechanics of Solids},
  shortjournal = {Mathematics and Mechanics of Solids},
  volume = {27},
  number = {12},
  pages = {2715--2750},
  issn = {1081-2865, 1741-3028},
  doi = {10.1177/10812865221078822},
  url = {https://journals.sagepub.com/doi/10.1177/10812865221078822},
}

@article{Favrie2017a,
  title = {A Rapid Numerical Method for Solving {{Serre-Green-Naghdi}} Equations Describing Long Free Surface Gravity Waves},
  author = {Favrie, N. and Gavrilyuk, S.},
  date = {2017},
  journaltitle = {Nonlinearity},
  volume = {30},
  number = {7},
  pages = {2718--2736},
  publisher = {IOP Publishing},
  issn = {13616544},
  doi = {10.1088/1361-6544/aa712d},
}

@article{Lombard2019,
  title = {Simulating Transient Wave Phenomena in Acoustic Metamaterials Using Auxiliary Fields},
  author = {Lombard, Bruno and Bellis, C and Lombard, B},
  date = {2019},
  journaltitle = {Wave Motion},
  volume = {86},
  pages = {175--194},
  publisher = {Elsevier B.V.},
  issn = {01652125},
  doi = {10.1016/j.wavemoti.2019.01.010},
  url = {https://doi.org/10.1016/j.wavemoti.2019.01.010},
}

@article{Dhaouadi2022a,
  title = {Hyperbolic Relaxation Models for Thin Films down an Inclined Plane},
  author = {Dhaouadi, Firas and Gavrilyuk, Sergey and Vila, Jean-paul},
  date = {2022},
  journaltitle = {Applied Mathematics and Computation},
  volume = {433},
  pages = {127378},
  publisher = {Elsevier Inc.},
  issn = {0096-3003},
  doi = {10.1016/j.amc.2022.127378},
  url = {https://doi.org/10.1016/j.amc.2022.127378},
}

@article{SHTC_surfacetension2025,
  title = {Nonequilibrium Model for Compressible Two-Phase Two-Pressure Flows with Surface Tension},
  author = {Peshkov, Ilya and Romenski, Evgeniy and Pavelka, Michal},
  date = {2025-09},
  journaltitle = {Continuum Mechanics and Thermodynamics},
  shortjournal = {Continuum Mech. Thermodyn.},
  volume = {37},
  number = {5},
  publisher = {{Springer Science and Business Media LLC}},
  issn = {0935-1175, 1432-0959},
  doi = {10.1007/s00161-025-01403-x},
  url = {https://link.springer.com/10.1007/s00161-025-01403-x},
}

@article{Dhaouadi2019,
  title = {Extended {{Lagrangian}} Approach for the Defocusing Nonlinear {{Schrödinger}} Equation},
  author = {Dhaouadi, Firas and Favrie, Nicolas and Gavrilyuk, Sergey},
  date = {2019-04},
  journaltitle = {Studies in Applied Mathematics},
  shortjournal = {Stud Appl Math},
  volume = {142},
  number = {3},
  pages = {336--358},
  issn = {0022-2526, 1467-9590},
  doi = {10.1111/sapm.12238},
  url = {https://onlinelibrary.wiley.com/doi/10.1111/sapm.12238},
}

@article{Mazaheri2016,
  title = {A First-Order Hyperbolic System Approach for Dispersion},
  author = {Mazaheri, Alireza and Ricchiuto, Mario and Nishikawa, Hiroaki},
  date = {2016-09},
  journaltitle = {Journal of Computational Physics},
  volume = {321},
  pages = {593--605},
  issn = {00219991},
  doi = {10.1016/j.jcp.2016.06.001},
  url = {http://www.sciencedirect.com/science/article/pii/S0021999116302261},
}

@article{Romenski2011,
  title = {On Modeling the Frequency Transformation Effect in Elastic Waves},
  author = {Romenski, E I and Sadykov, A D},
  year = {2011-04-31},
  journaltitle = {Journal of Applied and Industrial Mathematics},
  volume = {5},
  number = {2},
  pages = {282--289},
  publisher = {Pleiades Publishing, Ltd},
  issn = {1990-4789},
  doi = {10.1134/S1990478911020153},
  url = {http://link.springer.com/10.1134/S1990478911020153},
}

@article{Gavrilyuk2005,
  title = {Acoustic Properties of a Two-Fluid Compressible Mixture with Micro-Inertia},
  author = {Gavrilyuk, S. L.},
  date = {2005},
  journaltitle = {European Journal of Mechanics, B/Fluids},
  volume = {24},
  number = {3},
  pages = {397--406},
  issn = {09977546},
  doi = {10.1016/j.euromechflu.2004.09.001},
}

@incollection{GodRom1995,
  title = {Thermodynamics, Conservation Laws and Symmetric Forms of Differential Equations in Mechanics of Continuous Media},
  booktitle = {Computational {{Fluid Dynamics Review}} 1995},
  author = {Godunov, S.K. and Romensky, E.I.},
  date = {1995},
  volume = {95},
  pages = {19--31},
  publisher = {John Wiley, NY},
  doi = {10.1142/7799},
 }

@inproceedings{GodRom1996,
  title = {Symmetric Forms of Thermodynamically Compatible Systems of Conservation Laws in Continuum Mechanics},
  booktitle = {{{ECCOMAS Conference}} on Numerical Methods in Engineering},
  author = {Godunov, S K and Romensky, E I},
  date = {1996},
  pages = {54--57},
}

@book{Serre2007,
  title = {Multi-dimensional hyperbolic partial differential equations},
  author = {Benzoni-Gavage, Sylvie and Serre, Denis},
  date = {2006-11-23},
  volume = {325},
  publisher = {Oxford University Press},
  location = {Berlin, Heidelberg},
  doi = {10.1093/acprof:oso/9780199211234.001.0001},
  url = {https://oxford.universitypressscholarship.com/view/10.1093/acprof:oso/9780199211234.001.0001/acprof-9780199211234},
  isbn = {978-0-19-921123-4},
}

@article{Kato1975,
  title = {The {{Cauchy}} Problem for Quasi-Linear Symmetric Hyperbolic Systems},
  author = {Kato, Tosio},
  date = {1975},
  journaltitle = {Archive for Rational Mechanics and Analysis},
  volume = {58},
  number = {3},
  pages = {181--205},
  publisher = {Springer-Verlag},
  issn = {0003-9527},
  doi = {10.1007/BF00280740},
  url = {http://link.springer.com/10.1007/BF00280740},
}

@incollection{Rom2001,
  title = {Thermodynamics and {{Hyperbolic Systems}} of {{Balance Laws}} in {{Continuum Mechanics}}},
  booktitle = {Godunov {{Methods}}},
  author = {Romensky, Evgeniy I},
  editor = {Toro, E. F.},
  date = {2001},
  pages = {745--761},
  publisher = {Springer US},
  location = {New York, NY},
  doi = {10.1007/978-1-4615-0663-8_75},
  url = {http://www.springer.com/gp/book/9780306466014},
  isbn = {978-1-4613-5183-2},
}

@article{Rom1998,
  title = {Hyperbolic Systems of Thermodynamically Compatible Conservation Laws in Continuum Mechanics},
  author = {Romensky, E I},
  date = {1998},
  journaltitle = {Mathematical and computer modelling},
  volume = {28},
  number = {10},
  pages = {115--130},
  doi = {10.1016/S0895-7177(98)00159-9},
  url = {https://www.sciencedirect.com/science/article/pii/S0895717798001599},
}

@book{GodRom2003,
  title = {Elements of {{Continuum Mechanics}} and {{Conservation Laws}}},
  author = {Godunov, Sergei K and Romenskii, Evgenii I},
  date = {2003},
  publisher = {Springer US},
  location = {Boston, MA},
  doi = {10.1007/978-1-4757-5117-8},
  url = {http://link.springer.com/10.1007/978-1-4757-5117-8},
  isbn = {978-1-4419-3399-7},
}

@article{God1961,
  title = {An Interesting Class of Quasilinear Systems},
  author = {Godunov, S.K. K},
  date = {1961},
  journaltitle = {Dokl. Akad. Nauk SSSR},
  volume = {139(3)},
  number = {3},
  pages = {521--523},
}

@software{MATLAB,
year = {2025},
author = {The MathWorks Inc.},
title = {MATLAB version: 25.2.0.3042426 (R2025b)},
publisher = {The MathWorks Inc.},
address = {Natick, Massachusetts, United States},
url = {https://www.mathworks.com}
}

@article{SHTC-GENERIC,
  title = {Continuum Mechanics and Thermodynamics in the {{Hamilton}} and the {{Godunov-type}} Formulations},
  author = {Peshkov, Ilya and Pavelka, Michal and Romenski, Evgeniy and Grmela, Miroslav},
  date = {2018-11-18},
  journaltitle = {Continuum Mechanics and Thermodynamics},
  volume = {30},
  number = {6},
  eprint = {1710.00058},
  eprinttype = {arXiv},
  pages = {1343--1378},
  issn = {0935-1175},
  doi = {10.1007/s00161-018-0621-2},
  url = {http://link.springer.com/10.1007/s00161-018-0621-2},
}

@article{Acharya2022,
  title = {An Action for Nonlinear Dislocation Dynamics},
  author = {Acharya, Amit},
  date = {2022-04},
  journaltitle = {Journal of the Mechanics and Physics of Solids},
  volume = {161},
  pages = {104811},
  publisher = {Elsevier BV},
  issn = {0022-5096},
  doi = {10.1016/j.jmps.2022.104811},
  url = {https://linkinghub.elsevier.com/retrieve/pii/S0022509622000291},
}

@article{Dzyaloshinskii1980,
  title = {Poisson brackets in condensed matter physics},
  author = {Dzyaloshinskii, I. E. and Volovick, G. E.},
  date = {1980},
  journaltitle = {Annals of Physics},
  volume = {125},
  number = {1},
  pages = {67--97},
  issn = {1096035X},
  doi = {10.1016/0003-4916(80)90119-0},
}

@article{clayton2006,
  title = {Modeling Dislocations and Disclinations with Finite Micropolar Elastoplasticity},
  author = {Clayton, J.D. and McDowell, D.L. and Bammann, D.J.},
  date = {2006-02},
  journaltitle = {International Journal of Plasticity},
  shortjournal = {International Journal of Plasticity},
  volume = {22},
  number = {2},
  pages = {210--256},
  issn = {07496419},
  doi = {10.1016/j.ijplas.2004.12.001},
  url = {https://linkinghub.elsevier.com/retrieve/pii/S0749641905000409},
}

@article{Katanaev2005,
  title = {Geometric Theory of Defects},
  author = {Katanaev, Mikhail O},
  date = {2005-07-31},
  journaltitle = {Physics-Uspekhi},
  shortjournal = {Phys.-Usp.},
  volume = {48},
  number = {7},
  pages = {675--701},
  publisher = {Uspekhi Fizicheskikh Nauk (UFN) Journal},
  issn = {1063-7869, 1468-4780},
  doi = {10.1070/pu2005v048n07abeh002027},
  url = {https://ufn.ru/en/articles/2005/7/b/},
}

@incollection{Eringen1968,
  title = {Mechanics of Micromorphic Continua},
  booktitle = {Mechanics of Generalized Continua},
  author = {Eringen, A. C.},
  editor = {Kröner, E},
  date = {1968},
  pages = {18--35},
  publisher = {Springer Berlin Heidelberg},
  location = {Berlin, Heidelberg},
  doi = {10.1007/978-3-662-30257-6_2},
  url = {http://link.springer.com/10.1007/978-3-662-30257-6_2},
}

@article{Peshkov2025,
  title = {First-Order Hyperbolic Formulation of the Teleparallel Gravity Theory},
  author = {Peshkov, Ilya and Olivares, Héctor and Romenski, Evgeniy},
  date = {2025-10-27},
  journaltitle = {Physical Review D},
  shortjournal = {Phys. Rev. D},
  volume = {112},
  number = {8},
  eprint = {2211.13659},
  eprinttype = {arXiv},
  pages = {084070},
  issn = {2470-0010, 2470-0029},
  doi = {10.1103/jb23-c4pd},
  url = {https://link.aps.org/doi/10.1103/jb23-c4pd},
}

@misc{Kleinert1989,
  title = {Gauge Fields in Condensed Matter},
  author = {Kleinert, H},
  doi = {10.1142/0356},
  publisher={World Scientific},
  url = {https://doi.org/10.1142/0356},
}

@article{Mindlin1964,
  title = {Micro-Structure in Linear Elasticity},
  author = {Mindlin, R. D.},
  date = {1964-01},
  journaltitle = {Archive for Rational Mechanics and Analysis},
  volume = {16},
  number = {1},
  pages = {51--78},
  issn = {0003-9527},
  doi = {10.1007/BF00248490},
  url = {http://link.springer.com/10.1007/BF00248490},
  isbn = {0003-9527},
}

@article{NguyenLeMarrec2022,
  title = {On Tangent Geometry and Generalised Continuum with Defects},
  author = {Nguyen, Van Hoi and Casale, Guy and Le Marrec, Loïc},
  date = {2022-07-27},
  journaltitle = {Mathematics and Mechanics of Solids},
  volume = {27},
  number = {7},
  pages = {1255--1283},
  issn = {1081-2865},
  doi = {10.1177/10812865211059222},
  url = {http://journals.sagepub.com/doi/10.1177/10812865211059222},
}

@article{CRESPO2025a,
  title = {Two-Scale Geometric Modelling for Defective Media: {{An}} Approach Using Fibre Bundles},
  shorttitle = {Two-Scale Geometric Modelling for Defective Media},
  author = {Crespo, Mewen and Casale, Guy and Le Marrec, Loïc},
  date = {2025-06-17},
  journaltitle = {Mathematics and Mechanics of Solids},
  shortjournal = {Mathematics and Mechanics of Solids},
  pages = {10812865241312135},
  issn = {1081-2865, 1741-3028},
  doi = {10.1177/10812865241312135},
  url = {https://journals.sagepub.com/doi/10.1177/10812865241312135},
  urldate = {2025-06-24},
  langid = {english},
}

@book{LychevKoifman2018,
  title = {Geometry of Incompatible Deformations},
  author = {Lychev, Sergey and Koifman, Konstantin},
  date = {2018-11-05},
  publisher = {De Gruyter},
  location = {Berlin, Boston},
  doi = {10.1515/9783110563214},
  url = {https://www.degruyter.com/doi/10.1515/9783110563214},
  isbn = {978-3-11-056321-4},
}

@article{Hehl2007,
  title = {Elie Cartan's torsion in geometry and in field theory, an essay},
  author = {Hehl, Friedrich W. and Obukhov, Yuri N.},
  date = {2007-11-09},
  journaltitle = {Annales de la Fondation Louis de Broglie},
  volume = {32},
  number = {2--3},
  eprint = {0711.1535},
  eprinttype = {arXiv},
  pages = {157--194},
  issn = {01824295},
  url = {http://arxiv.org/abs/0711.1535},
  urldate = {2018-01-14},
  langid = {italian},
}

@article{MALYSHEV2007,
  title = {The {{Einsteinian}}{{{\mkbibemph{T}}}}(3)-Gauge Approach and the Stress Tensor of the Screw Dislocation in the Second Order: Avoiding the Cut-off at the Core},
  shorttitle = {The {{Einsteinian}}{{{\mkbibemph{T}}}}(3)-Gauge Approach and the Stress Tensor of the Screw Dislocation in the Second Order},
  author = {Malyshev, C},
  date = {2007-08-24},
  journaltitle = {Journal of Physics A: Mathematical and Theoretical},
  shortjournal = {J. Phys. A: Math. Theor.},
  volume = {40},
  number = {34},
  pages = {10657--10684},
  publisher = {IOP Publishing},
  issn = {1751-8113, 1751-8121},
  doi = {10.1088/1751-8113/40/34/019},
  url = {https://iopscience.iop.org/article/10.1088/1751-8113/40/34/019},
  urldate = {2025-07-07},
  langid = {english},
}

@article{lazar2000,
  title = {Dislocation theory as a 3-dimensional translation gauge theory},
  author = {Lazar, Markus},
  date = {2000},
  journaltitle = {Annalen der Physik (Leipzig)},
  volume = {9},
  number = {6},
  eprint = {cond-mat/0006280},
  eprinttype = {arXiv},
  pages = {461--473},
  issn = {00033804},
  doi = {10.1002/1521-3889(200006)9:6<461::AID-ANDP461>3.0.CO;2-B},
}

@article{Lazar2002a,
  title = {An Elastoplastic Theory of Dislocations as a Physical Field Theory with Torsion},
  author = {Lazar, Markus},
  date = {2002},
  journaltitle = {Journal of Physics A: Mathematical and General},
  volume = {35},
  number = {8},
  eprint = {cond-mat/0105270},
  eprinttype = {arXiv},
  pages = {1983--2004},
  issn = {0305-4470},
  doi = {10.1088/0305-4470/35/8/313},
  url = {http://arxiv.org/abs/cond-mat/0105270%5Cnhttp://stacks.iop.org/0305-4470/35/i=8/a=313?key=crossref.d765d9274e940a6f3b2af670124171ca},
}

@article{Cartan1922,
  title = {Sur Une Généralisation de La Notion de Courbure de {{Riemann}} et Les Espaces à Torsion},
  author = {{Cartan, É}},
  date = {1922},
  journaltitle = {C R Acad Sci},
  volume = {174},
  pages = {593--595}
}

@article{Toupin1962,
  title = {Elastic Materials with Couple-Stresses},
  author = {Toupin, R. A.},
  date = {1962},
  journaltitle = {Archive for Rational Mechanics and Analysis},
  volume = {11},
  number = {1},
  pages = {385--414},
  issn = {0003-9527},
  doi = {10.1007/BF00253945},
  url = {http://link.springer.com/10.1007/BF00253945},
}

@article{Madeo2015a,
  title = {Wave Propagation in Relaxed Micromorphic Continua: Modeling Metamaterials with Frequency Band-Gaps},
  author = {Madeo, A. and Neff, P. and Ghiba, I. D. and Placidi, L. and Rosi, G.},
  date = {2015-09-19},
  journaltitle = {Continuum Mechanics and Thermodynamics},
  volume = {27},
  number = {4--5},
  pages = {551--570},
  publisher = {Springer Berlin Heidelberg},
  issn = {0935-1175},
  doi = {10.1007/s00161-013-0329-2},
  url = {http://link.springer.com/10.1007/s00161-013-0329-2},
  urldate = {2019-02-13},
}

@article{Bilby1955,
  title = {Continuous {{Distributions}} of {{Dislocations}}: {{A New Application}} of the {{Methods}} of {{Non-Riemannian Geometry}}},
  author = {Bilby, B. A. and Bullough, R. and Smith, E.},
  date = {1955-08-22},
  journaltitle = {Proceedings of the Royal Society A: Mathematical, Physical and Engineering Sciences},
  volume = {231},
  number = {1185},
  pages = {263--273},
  publisher = {The Royal Society},
  issn = {1364-5021},
  doi = {10.1098/rspa.1955.0171},
  url = {http://rspa.royalsocietypublishing.org/cgi/doi/10.1098/rspa.1955.0171},
  urldate = {2018-02-09},
}

@article{Kosevich1965,
  title = {Dynamical Theory of Dislocation},
  author = {Kosevich, M.A.},
  date = {1965},
  journaltitle = {Sov. Phys. Usp},
  volume = {7},
  pages = {837},
}

@incollection{Kroner1963,
  title = {The {{Dislocation}} as a {{Fundamental New Concept}} in {{Continuum Mechanics}}},
  booktitle = {Materials {{Science Research}}},
  author = {Kröner, Ekkehart},
  editor = {Stadelmaier, H.H. and Austin, W.W.},
  date = {1963},
  pages = {281--290},
  publisher = {Springer US},
  location = {Boston, MA},
  doi = {10.1007/978-1-4899-5537-1_14},
  url = {http://link.springer.com/10.1007/978-1-4899-5537-1_14},
  urldate = {2018-02-09},
}

@article{Nye1953,
  title = {Some Geometrical Relations in Dislocated Crystals},
  author = {Nye, J.F},
  date = {1953-03-01},
  journaltitle = {Acta Metallurgica},
  volume = {1},
  number = {2},
  pages = {153--162},
  publisher = {Pergamon},
  issn = {00016160},
  doi = {10.1016/0001-6160(53)90054-6},
  url = {https://www.sciencedirect.com/science/article/pii/0001616053900546},
  urldate = {2018-02-09},
}

@article{Cosserat1909,
  title = {Théorie Des Corps Déformables},
  author = {Cosserat, Eugene and Cosserat, François},
  date = {1909},
  publisher = {A. Hermann et fils}
}

@article{Eringen1964a,
  title = {Nonlinear Theory of Simple Micro-Elastic Solids—{{I}}},
  author = {Eringen, A.Cemal and Suhubi, E.S.},
  date = {1964-05},
  journaltitle = {International Journal of Engineering Science},
  volume = {2},
  number = {2},
  pages = {189--203},
  issn = {00207225},
  doi = {10.1016/0020-7225(64)90004-7},
  url = {https://linkinghub.elsevier.com/retrieve/pii/0020722564900047},
}

@article{Cummer2016,
author = {Cummer, Steven A. and Christensen, Johan and Al{\`{u}}, Andrea},
doi = {10.1038/natrevmats.2016.1},
issn = {2058-8437},
journal = {Nature Reviews Materials},
month = {mar},
number = {3},
pages = {16001},
title = {{Controlling sound with acoustic metamaterials}},
url = {http://www.nature.com/articles/natrevmats20161},
volume = {1},
year = {2016}
}

@article{Nabarro1951,
  title = {The {{Interaction}} of {{Screw Dislocations}} and {{Sound Waves}}},
  author = {Nabarro, F. R. N.},
  date = {1951},
  journaltitle = {Proceedings of the Royal Society of London. Series A, Mathematical and Physical Sciences},
  volume = {209},
  number = {1097},
  eprint = {98896},
  eprinttype = {jstor},
  pages = {278--290},
  url = {http://www.jstor.org/stable/98896},
}

@article{PRD-Torsion2019,
	archivePrefix = {arXiv},
	arxivId = {1810.03761},
	author = {Peshkov, Ilya and Romenski, Evgeniy and Dumbser, Michael},
	doi = {10.1007/s00161-019-00770-6},
	eprint = {1810.03761},
	issn = {0935-1175},
	journal = {Continuum Mechanics and Thermodynamics},
	month = {sep},
	number = {5},
	pages = {1517--1541},
	publisher = {Springer Berlin Heidelberg},
	title = {{Continuum mechanics with torsion}},
	url = {http://link.springer.com/10.1007/s00161-019-00770-6
	http://arxiv.org/abs/1810.03761},
	volume = {31},
	year = {2019}
}

@book{Eringen1991,
  title = {Electrodynamics of Continua I},
  author = {Eringen, A. Cemal and Maugin, Gérard A.},
  date = {1990},
  volume = {58},
  number = {2},
  publisher = {Springer New York},
  location = {New York, NY},
  issn = {00218936},
  doi = {10.1007/978-1-4612-3226-1},
  url = {http://appliedmechanics.asmedigitalcollection.asme.org/article.aspx?articleid=1410431},
  isbn = {978-1-4612-7923-5},
}

@book{EringenMauginII,
  title = {Electrodynamics of Continua II},
  author = {Eringen, A. C. and Maugin, G. A.},
  date = {1990},
  publisher = {Springer New York},
  location = {New York, NY},
  doi = {10.1007/978-1-4612-3236-0},
  url = {http://link.springer.com/10.1007/978-1-4612-3236-0},
  isbn = {978-1-4612-7928-0},
}
